\documentstyle[psfig,twocolumn,aps,graphicx,eucal]{revtex}       
\tighten
\begin{document}

\setlength{\unitlength}{1cm}

\draft

\title{\bf Perturbation Theory by Flow Equations: 
Dimerized and Frustrated $S=1/2$ Chain}

\author{Christian Knetter\thanks{
e-mail: ck@thp.uni-koeln.de
\hspace*{\fill} {\protect\linebreak}
internet: www.thp.uni-koeln.de/\~{}ck/} and G\"otz S. 
Uhrig\thanks{e-mail: gu@thp.uni-koeln.de
\hspace*{\fill} {\protect\linebreak}
internet: www.thp.uni-koeln.de/\~{}gu/}}

\address{Institut f\"ur Theoretische Physik, Universit\"at zu
  K\"oln, Z\"ulpicher Str. 77, K\"oln 50937, Germany \\
  {\rm(\today)} }

\maketitle
\begin{abstract}
The flow equation method (Wegner 1994) is used as continuous
unitary transformation to construct perturbatively
effective Hamiltonians. The method is illustrated in detail
for dimerized and frustrated antiferromagnetic $S=1/2$ chains.
 The effective Hamiltonians
conserve the number of elementary excitations which are $S=1$ magnons for the
dimerized chains. The sectors of different
number of excitations are clearly separated.
Easy-to-use results for the gap, the dispersion and the ground 
state energies of the chains are provided.
\end{abstract}


\narrowtext
\section{Introduction}
Perturbation theory is one of the most important and most versatile
tools for problems which are not exactly solvable.
Various methods depending on the problem under study
have been invented and used. Due to the enormous increase
in computer capacity it is a very interesting task to
use algebraic programmes to perform perturbative
calculations.

The aim of the present work is to propose a general 
perturbation scheme which splits naturally into two
subsequent steps. Both these steps can be implemented
in a direct manner on the computer. The first step
is not model specific. It relies only on two prerequisites.
\begin{itemize}
\item[(i)]
The unperturbed Hamiltonian $H_0$ must have an equidistant
spectrum bounded from below. Without loss of generality
we may assume that $E_i = i$ for $i\in \{0,1,2,3,\ldots\}$.
We say that $i$ denotes the number of energy quanta in
the system. By $U_i$ the corresponding subspaces are denoted.
\item[(ii)]
The perturbing
Hamiltonian $H_{\rm S}$ links subspaces $U_i$ and $U_j$
only if $|i-j|$ is bounded from above, i.e. there is a
number $N>0$ such that $H_{\rm S}$ can be written as
$H_{\rm S} = \sum_{n=-N}^N T_n$ where $T_n$ increments 
(or decrements, if $n<0$) the number of energy quanta by
$n$
\begin{equation}
\label{outset0}
[ H_0,T_n] = n T_n\ .
\end{equation}
\end{itemize}
 Thus the full problem reads
\begin{equation}
\label{outset1}
H = H_0 + \lambda\sum_{n=-N}^N T_n
\end{equation}
where $\lambda$ is the perturbation parameter supposed to 
be small $\lambda < 1$.
In this work we will restrict to $N=2$.
The first step consists in finding a systematic mapping
of the problem in Eq.~(\ref{outset1}) to an effective one given by
a Hamiltonian
$H_{\rm eff}$ which conserves the number of energy quanta.

The second step is the model specific one. It consists in
the actual calculation of $H_{\rm eff}$ for a given number
of energy quanta.

To illustrate the abstract ideas formulated above we will
use the frustrated and dimerized $S=1/2$ chain given by (j counts the sites)
\begin{equation}
\label{hamil1}
H = J_0\sum_{j=0}^{L}\left[ (1+(-1)^j \delta) {\bf S}_j{\bf S}_{j+1}
+\alpha_0{\bf S}_{j-1}{\bf S}_{j+1}\right] \ ,
\end{equation}
where $L$ is the number of sites. A situation consistent with
Eq.~(\ref{outset1}) is found
for strong dimerization. Hence we rewrite Hamiltonian (\ref{hamil1}) as 
(subscript $i$ counts the dimers)
\begin{eqnarray}
\nonumber
H &=& J\sum_{i=0}^{\frac{L}{2}-1} \left[ {\bf S}_{2i}{\bf S}_{2i+1}+ \lambda
{\bf S}_{2i}{\bf S}_{2i-1} +\right.\qquad\qquad\qquad\\\label{hamil2}
&& \qquad\qquad \qquad\lambda\left.
\alpha({\bf S}_{2i}{\bf S}_{2i-2}+{\bf S}_{2i-1}{\bf S}_{2i+1})\right] 
\end{eqnarray}
with 
\begin{mathletters}
\label{substitution}
\begin{eqnarray}
J &=& J_0(1+\delta)\\
\lambda  &=& (1-\delta)/(1+\delta)\\
\alpha &=& \alpha_0/(1-\delta) \ .
\end{eqnarray}
\end{mathletters}
The unperturbed part $H_0$ 
(up to a trivial constant $3L/8$)
and the perturbing part $H_{\rm S}$ are then 
\begin{mathletters}
\label{H_ges}
\begin{eqnarray}
\label{H0_1}
H_0 &=& \sum_i \left[ {\bf S}_{2i}{\bf S}_{2i+1}+3/4 \right]\\\label{HS_1}
H_{\rm S} &=& \sum_i \left[{\bf S}_{2i}{\bf S}_{2i-1} +
\alpha\left({\bf S}_{2i}{\bf S}_{2i-2}+{\bf S}_{2i-1}{\bf S}_{2i+1}
\right)\right] \ ,
\end{eqnarray}
\end{mathletters}
where we measure implicitly all energies in units of $J$.
The ground state of $H_0$ is the product of singlets on the
dimers, i.e. the bonds $(2i,2i+1)$.
The energy quanta are here the excited dimers, namely the local triplets.
The number of triplets classifies the degenerate energy eigen spaces
of the unperturbed problem.

Besides the purpose to serve as an example for perturbation
by flow equations the frustrated and dimerized chain is 
of considerable physical interest itself. 
Ideal spin-Peierls systems are one-dimensional spin systems
which are coupled to the lattice. At low enough temperatures
they dimerize since this dimerization leads to a gain in magnetic
energy $\propto \delta^{4/3}$ which overcompensates the loss in
elastic energy $\propto \delta^{2}$, see e.g. \cite{bray83} and refs.~therein.
So spin-Peierls systems provide dimerized spin chains in a natural way. 
The first inorganic spin-Peierls substance CuGeO$_3$
 in particular provides the example of a frustrated and dimerized
spin chain since there is much evidence that a certain amount of
frustration is present in this substance \cite{riera95,casti95,fabri98a}.
Other examples are strongly anisotropic substances where the
dimerization is built-in in the chemical structure. Examples
are Cu$_2$(C$_5$H$_{12}$N$_2$)$_2$Cl$_4$ \cite{chabo97}
and (VO)$_2$P$_2$O$_7$ \cite{garre97a}.
Of course,  the real substance mostly display also some
additional two- or three-dimensional coupling. But the approach
we present here is suited to tackle even these systems, see e.g.
Ref. \cite{uhrig98c}.

One might argue that exact diagonalization or quantum Monte Carlo
approaches are better suited to calculate  dispersions $\omega(k)$
or similar quantities in $d=1$. These methods, however, yield
only the result for the chosen parameter set. The perturbative results,
however, will be obtained as polynomials in the weak bond coupling
$\lambda$ and the frustration $\alpha$. Thus, once computed, anyone
can use the perturbative results easily to 
fit measured or otherwise obtained data.
Thereby an extremely fast method for the determination of
coupling constants is provided. Of course, the perturbative approach
can be applied only for $\lambda\le1$ where the equal sign represents
the worst case. For $\lambda > 1$ the perturbative approach
breaks down.

The work is organized as follows. In the next section
we extend the approach of Stein \cite{stein97} who did 
a calculation for $N=1$ up to fifth order 
to $N=2$ and up to tenth order. 
The work of Stein improved earlier calculations \cite{macdo88}
which generated more intermediate terms (see Ref.~\cite{stein97}
for discussion).
The flow equation transformation
which is used by Stein and by us was introduced by
Wegner five years ago \cite{wegne94}.
In Sect. III we illustrate our method by applying
it to a one-dimensional Heisenberg antiferromagnetic
$S=1/2$ chain. The effective one-triplet Hamiltonian
is computed. The ground state energy, gaps and dispersion
relations are discussed in Sect.~IV. Summary and outlook
conclude the main part of our work.

\section{Perturbation by Flow Equations}

All what is presented in this section is based only on
the fact that the initial problem has the form (\ref{outset1})
fulfilling the requirements (i) and (ii) with $N=2$.

The general idea behind the flow equation approach introduced
by Wegner \cite{wegne94} is to perform a {\em continuous}
unitary transformation which makes the problem more easily
tractable. Mostly, one tries to make the problem ``more
diagonal''. In our case we will achieve a block diagonal form.
A broad field of application is to identify certain quasi-particles
for which an effective Hamiltonian can be found. Here, flow equations
can be used to implement a renormalization of a given problem on the
Hamiltonian level, not only  on the level of certain observables or couplings
\cite{wegne94,kehre96a,kehre96b}. Analogous ideas were 
suggested parallely by G{\l}azek
and Wilson in the form of similarity transformations \cite{glaze93}.

In the present work we do not focus on the renormalization properties of
the flow equation approach. Following Stein \cite{stein97}, we use them to
 implement in a systematic way a continuous unitary transformation which maps 
the perturbed system onto the unperturbed one which is easy to understand.

\subsection{General Formalism}
According to the original idea a running variable $\ell$ is 
introduced which parameterizes the continuously evolving 
Hamiltonian $H(\ell)$. The starting operator is the bare
Hamiltonian;
the operator at infinity is the desired effective Hamiltonian
\begin{mathletters}
\begin{eqnarray}
\label{startcond}
H(0) &=& H_0+\lambda (T_{-2}+T_{-1}+T_{0}+T_{1}+T_{2})\\
H(\infty) &=& H_{\rm eff}\ .
\end{eqnarray}
\end{mathletters}
 The unitary evolution is engendered by its
antihermitean infinitesimal generator $\eta(\ell)$
\begin{equation}
\label{flow}
\frac{dH(\ell)}{d\ell} = [\eta(\ell), H(\ell)]\ .
\end{equation}
Applying naively Wegner's choice for the generator
 $\eta(\ell) =[H_0,H(\ell)]$ the resulting differential equations
quickly become very messy since the band block diagonal structure
of the original problem is lost.
By ``band block diagonal'' we mean the fact that $N$ has a finite value
which does not change in the course of the flow $\ell\to\infty$. 

We will choose a slightly
different infinitesimal generator which allows to
keep the band block diagonal structure of the original problem, i.e.
the parameter $N$ stays 2 for all values of $\ell$. The general
Hamiltonian $H(\ell)$ can be written as
\begin{equation}
\label{hamdef}
H(\ell) = H_0+\lambda \Theta(\ell)
\end{equation}
and the operator $\Theta(\ell)$ links only subspaces $U_i$ and
$U_j$ with $|i-j|\leq 2$.

The most general form of $\Theta(\ell)$ is
\begin{equation}
\label{thetadef}
\Theta(\ell) = \sum_{k=1}^{\infty}\lambda^{k-1}
\sum_{ |\underline{m}|=k }
F(\ell;\underline{m}) T(\underline{m})
\end{equation}
where $F(\ell;\underline{m})$ are real-valued functions
for which we will derive nonlinear, but recursive, differential equations
below. The other symbols are
\begin{mathletters}
\begin{eqnarray}
\underline{m} &=& (m_1,m_2,m_3,\ldots,m_k)\quad\mbox{with}\\\label{m_menge}
 m_i &\in& \{ 0,\pm1,\pm2\}\\\label{m_betrag}
|\underline{m}| &=& k \\
T(\underline{m}) &=& T_{m_1}T_{m_2}T_{m_3}\ldots T_{m_k}\\
M(\underline{m}) &=& \sum_{i=1}^k m_i \ .
\end{eqnarray}
\end{mathletters}
The vector $\underline{m}$ together with the product $T(\underline{m})$
encode all possible products of the incrementing (decrementing)
operators $T_n$ as defined in Eqs. (\ref{outset0},\ref{outset1}).
The infinitesimal generator of our choice reads
\begin{equation}
\label{infgen}
\eta(\ell) = \sum_{k=1}^\infty\lambda^k
\sum_{|\underline{m}|=k} {\rm sgn}(M(\underline{m}))
F(\ell;\underline{m}) T(\underline{m})\ .
\end{equation}
This choice is very similar
to the infinitesimal generator Mielke proposes
for band matrices \cite{mielk98}. An adapted version of
his proof that such an $\eta$ leads to (block) diagonality can be found
in appendix \ref{app-proof}. For the purposes of the present perturbative
approach the general proof can be replaced by the observation
that the transformation can be performed successfully to all finite
orders. This will be shown below.

A short computation shows that substituting ${\rm sgn}(M(\underline{m}))$
in Eq.~(\ref{infgen}) by $M(\underline{m})$ would correspond to the first
 choice $\eta(\ell) =[H_0,H(\ell)]$. This relies on 
\begin{equation}
[H_0,T(\underline{m})] = M(\underline{m})T(\underline{m})
\label{kommut-m}
\end{equation}
which is a straightforward generalization of Eq.~(\ref{outset0}). 

Insertion of the ans\"atze (\ref{thetadef}) and (\ref{infgen}) 
into Eq.~(\ref{flow}) leads to
\begin{eqnarray}\nonumber
\lambda\frac{d\Theta}{d\ell} &=& \lambda[\eta(\ell),\Theta(\ell)]
-\\
&&  \sum_{k=1}^\infty \lambda^k
\sum_{|\underline{m}| = k} 
{\rm sgn}(M(\underline{m}))F(\ell;\underline{m}) [H_0,T(\underline{m})]\ .
\end{eqnarray}
Comparison of the coefficients for each term $T(\underline{m})$
yields then a differential equation for the functions 
$F(\ell;\underline{m})$
\begin{eqnarray}\label{dgl}
\frac{d}{d\ell}F(\ell;\underline{m}) &=& -|M(\underline{m})|  
F(\ell;\underline{m})+  \\\nonumber
&& \hspace{-19mm}\sum_{\{ \underline{m}_1,\underline{m}_2\}
=\underline{m}} \hspace{-4mm} \left[{\rm sgn}(M(\underline{m}_1)) -
{\rm sgn}(M(\underline{m}_2))  \right]
F(\ell;\underline{m}_1)
F(\ell;\underline{m}_2).
\end{eqnarray}
The summation condition $\{ \underline{m}_1,\underline{m}_2\}=\underline{m}$
means that one sums over all possible nontrivial breakups of $\underline{m}$
\begin{eqnarray}\nonumber
\underline{m}_1 =(m_1) &\ \mbox{and}\ & 
\underline{m}_2 =(m_2,\ldots,m_k)\\\nonumber
\underline{m}_1 =(m_1,m_2) &\ \mbox{and}\ & 
\underline{m}_2 =(m_3,\ldots,m_k)\\\nonumber
\underline{m}_1 =(m_1,m_2,m_3) &\ \mbox{and}\ & 
\underline{m}_2 =(m_4,\ldots,m_k)\\\nonumber
&\vdots&\\\nonumber
\underline{m}_1 =(m_1,\ldots,m_{k-2}) &\ \mbox{and}\ & 
\underline{m}_2 =(m_{k-1},m_k)\\
\underline{m}_1 =(m_1,\ldots,m_{k-1}) &\ \mbox{and}\ & 
\underline{m}_2 =(m_k)\ .\label{breakups}
\end{eqnarray}
This summation notation will also be used in the following.
The starting conditions follow from (\ref{startcond})
\begin{mathletters}
\label{startcond2}
\begin{eqnarray}
F(0;\underline{m}) &=& 1 \quad \mbox{for} \quad |\underline{m}|=1\\
F(0;\underline{m}) &=& 0 \quad \mbox{for} \quad |\underline{m}|>1\ .
\end{eqnarray}
\end{mathletters}

From Eqs.~(\ref{dgl},\ref{startcond2}) we can deduce a number
of relations by induction. First, we see that the functions
$F(\ell;\underline{m})$ are always real. Furthermore,
they obey the two symmetry relations
\begin{mathletters}
\label{symrel}
\begin{eqnarray}
F(\ell;-\underline{\overline{m}}) &=& F(\ell;\underline{m})\\[2mm]
F(\ell;-\underline{m}) &=& (-1)^{|\underline{m}|+1}F(\ell;\underline{m})
\end{eqnarray}
\end{mathletters}
where we use the notation 
\begin{equation}
\underline{\overline{m}} = (m_k,m_{k-1},\ldots,m_2,m_1)\ .
\end{equation}
The square bracket in Eq.~(\ref{dgl}) ensures that the sum vanishes
if $|M(\underline{m})|>2$
\begin{equation}
F(\ell;\underline{m}) = 0 \quad \mbox{for}\quad |M(\underline{m})|>2 \ .
\end{equation}
 For instance, a term generating three energy
quanta $M(\underline{m})=3$ could only be induced
from terms with $M(\underline{m}_1)=2$ and $M(\underline{m}_2)=1$
or vice-versa. But such combinations are suppressed by the square
bracket in Eq.~(\ref{dgl}). 
This observation is at the basis of the preservation of
the band block structure \cite{mielk98}. If we had chosen the
infinitesimal generator $\eta(\ell)$ in (\ref{infgen}) without the
signum as it would correspond to Wegner's original suggestion
$\eta(\ell) =[H_0,H(\ell)]$ the square bracket in Eq.~(\ref{dgl}) 
would read $[M(\underline{m}_1)-M(\underline{m}_2)]$ and hence the 
band structure of the couplings would be destroyed for $\ell>0$.

For the solution of Eq.~(\ref{dgl}) we observe that the first term
on the right hand side just generates an exponential prefactor
\begin{equation}
\label{fansatz}
F(\ell;\underline{m}) = \exp(-|M(\underline{m})|\ell)
f(\ell;\underline{m})\ .
\end{equation}
The rest of the equation (\ref{dgl}) is recursive and can thus be 
directly found by integration beginning from the starting conditions
\begin{eqnarray}\nonumber
\frac{d}{d\ell}f(\ell;\underline{m}) &=& 
 \hspace{-3mm}
\sum_{\{ \underline{m}_1,\underline{m}_2\}
=\underline{m}} \hspace{-5mm} 
e^{(|M(\underline{m})|-|M(\underline{m}_1)|-|M(\underline{m}_2)|)l}\cdot\\
&& \hspace{-14mm} \left[{\rm sgn}(M(\underline{m}_1)) -
{\rm sgn}(M(\underline{m}_2))  \right]
f(\ell;\underline{m}_1)f(\ell;\underline{m}_2) \ .
\label{rekdgl}
\end{eqnarray}
Note that $|M(\underline{m})|-|M(\underline{m}_1)|-|M(\underline{m}_2)|\leq0$
holds so that no exponential growth occurs in the 
$f(\ell;\underline{m})$.
Let us focus on the functional form of the functions 
$f(\ell;\underline{m})$. We state  by induction
that sums of terms $\ell^ie^{-2\mu\ell}$ occur for non-negative
integers $i$ and $\mu$. More precisely, we obtain
\begin{equation}
\label{fform}
f(\ell;\underline{m}) = \sum^{\Gamma(\underline{m})}_{\mu=0}
P_\mu(\ell;\underline{m}) e^{-2\mu\ell}
\end{equation}
where the degree of the polynomials $P_\mu(\ell;\underline{m})$
is always equal or less than $|\underline{m}|$ and the upper limit of the
sum $\Gamma(\underline{m})$ obeys
\begin{equation}
\Gamma(\underline{m})= \frac{1}{2} \left(-|M(\underline{m})|+
\sum_{i=1}^{|\underline{m}|} |m_i| \right)\ .
\end{equation}
In principle it is also possible to write down explicit recursion
relations for the polynomials, see for instance Ref. \cite{stein97} for $N=1$.
But they are of little clarity.
If the actual calculation is done by symbolic calculation it is
sufficient to retain that according to Eq.~(\ref{dgl}) or 
Eq.~(\ref{rekdgl}) expressions of the type
(\ref{fform}) have to be multiplied, added and integrated.
This is a straightforward task and
can be implemented in symbolic programmes.

The quantities we are finally interested in are the coefficients
of $H_{\rm eff}=H(\ell=\infty)$. From Eq.~(\ref{fansatz}) we know
that only terms with $M(\underline{m})=0$ will not vanish for
$\ell\to \infty$. This is exactly what we intended to achieve 
since we want $H_{\rm eff}$ to commute with $H_0$ (cf.~Eq.~
(\ref{kommut-m})) so that the
number of energy quanta (triplets in our example) becomes
a conserved quantity. Hence, we can write the final result as
\begin{mathletters}
\label{effhamilfin}
\begin{eqnarray}
H_{\rm eff} &=& H_0 +\sum_{k=1}^{\infty}\lambda^{k} 
\sum_{|\underline{m}|=k, M(\underline{m})=0} C(\underline{m}) 
T(\underline{m})\\
C(\underline{m}) &=& F(\infty;\underline{m})\ .
\end{eqnarray}
\end{mathletters}
Further details on the computation of the coefficients will be
given in the next subsection. Results for the $C(\underline{m})$
are presented in appendix \ref{app-coeff}.

\subsection{Computer Aided Evaluation}
We implemented the coefficient computation in C++ because of its high 
performance and its class concept which we used to encode the basic data 
elements. As an example let us
 consider a generic fourth order coefficient 
(i.e. $k=4$ ) $f(l;\underline{m})$ :
\begin{equation}
\label{fexample}
  f(l;(1,1,1,-2))=\frac{1}{2}l+\frac{1}{2}e^{-2l}-\frac{1}{8}e^{-4l}-
\frac{3}{8} \ .
\end{equation}
This can be stored as a list of basic data elements like
\begin{equation}
\label{basicdats}
 \frac{p}{q}l^ie^{-2\mu l}
\end{equation}
each containing four separate integers $p,q,i,\mu$. In fact, $p$ and $q$ 
can become very large. So they have to be stored in a multiprecision data 
type like long long int on some Unix systems. Still, all computations can 
be done very fast in integers and the results are rigourously exact.

Eq.~(\ref{rekdgl}) is essential in the computation of the $f(l;\underline{m})$.
 The basic idea is to build two loops. The outer loop controls the order 
starting at $k=2$ since the initial conditions are the values of the 
$f(l;\underline{m})$ in first order. The inner loop generates all possible 
$\underline{m}$ in the current order $k$. A single $m_i$ in 
$\underline{m}=(m_1,\ldots,m_i,\ldots,m_k)$ can take five different values,
see Eq.~ (\ref{m_menge}). Thus we can introduce the loop variable 
$n \in \{0,1,\ldots 5^k-1\}$ 
\begin{equation}
\label{5ersys}
  n=\sum_{i=0}^5 a_i5^i \ .
\end{equation}
The coefficients $a_i \in \{0,1,2,3,4 \}$ are mapped uniquely onto the set 
$\{-2,-1,0,1,2 \}$. Finally we retain those $\underline{m}$, with 
$|M(\underline{m})| \le 2$. In this way, Eq.~(\ref{rekdgl}) can be
 calculated for each $k$ and $\underline{m}$.

The calculation of a single $f(l;\underline{m})$ via Eq.~(\ref{rekdgl}) can be 
split in four steps. (i) One has to encapsulate Eq.~(\ref{rekdgl}) in 
yet another loop controlling all possible breakups of $\underline{m}$ 
(cf.~Eq.~(\ref{breakups})). (ii) The functions $f(l;\underline{m_1})$ and 
$f(l;\underline{m_2})$ known from calculations in lower orders have to be
 multiplied for each breakup of $\underline{m}$. (iii) One has to sum over all
 breakups. (iv) Finally, the result from steps (i) to (iii) has to be 
integrated.

Since $p$ and $q$ in the basic data elements 
(\ref{basicdats}) can become very large during addition and multiplication, 
both operations employ Euklid's algorithm to generate maximally canceled 
fractions ${p}/{q}$. To minimize memory usage these operations contain 
simplification subroutines based on the quick-sort algorithm. 
These subroutines sort according to 
increasing powers of $l$ and $e^{-2l}$. Simultaneously, addends are
 identified and added if they are of equal type.
The resulting $f(l;\underline{m})$ consists of linearly independent addends 
only. Due to the quick-sort algorithm the computation time as function of 
the number of addends $n$ is only of order $n\ln(n)$. 

The final integration can be done easily. The functions $f(l;\underline{m})$ 
 break down to basic data elements (\ref{basicdats}) so that ($\alpha>0$)
\begin{mathletters}
\begin{eqnarray}
  \label{implinta}
  \int_0^l dl' l'^i &=& \frac{1}{i+1}l^{i+1}\\
  \label{implintb}
  \int_0^l dl' l'^i e^{-\alpha l'} &=& 
\frac{i!}{\alpha}\cdot\left\{\frac{1}{\alpha^i}-
e^{-\alpha l}\sum_{j=0}^i\frac{l^j}{j! \alpha^{i-j}}\right\}
\end{eqnarray}
\end{mathletters}
achieves the integration.

To calculate the $C(\underline{m})$ one has to perform the 
$l \rightarrow \infty$ limit on those $f(l;\underline{m})$ for which 
$|M(\underline{m})|=0$. Note that for $|M(\underline{m})|=0$ one always has 
$\alpha>0$ since  $\alpha =|M(\underline{m})|-|M(\underline{m}_1)|
-|M(\underline{m}_2)|=0$  and $|M(\underline{m})|=0$ implies
the vanishing of $|M(\underline{m}_1)|$ and of $|M(\underline{m}_2)|$.
Hence the right hand side of Eq.~(\ref{rekdgl}) vanishes due to the 
square bracket containing the sign functions. So no such
$f(l;\underline{m})$ is generated. Thus the 
$C(\underline{m})$ can be calculated by Eq.~(\ref{implintb}) omitting the 
term proportional to $e^{-\alpha l}$ on the right hand side.

The symmetry relations  (\ref{symrel}) can be used as a check of the results.
The $C(\underline{m})$ are saved in a file together with the corresponding 
$\underline{m}$ for later usage. They are given up to order $k=6$ in appendix 
\ref{app-coeff}. We intend to provide them up to order $k=10$ in electronic
 form on our homepages on appearance of this article.

Unfortunately, Eq.~(\ref{rekdgl}) implies also a natural limitation of the 
computation. Because of its recursive nature, the $f(l;\underline{m})$ of all 
preceding orders have to be stored. They are needed to derive the 
$f(l;\underline{m})$ in the current order. This leads to an exponential memory 
increase. To calculate all $C(\underline{m})$ to order $k=10$ inclusively, 
we used about 30,000,000 basic data elements (\ref{basicdats})
 occupying about 1GB RAM. Because 
of the extensive memory use we employed a SUN Ultra Enterprise 10000 which the
 Regional Computing Center of the University of Cologne kindly placed at our 
disposal. The calculation took about 12h. 

\section{Application:\\ Dimerized and Frustrated $S=1/2$ Chain}
In this section we demonstrate how the knowledge of the $C(\underline{m})$
in the effective Hamiltonian $H_{\rm{eff}}$ (\ref{effhamilfin}) permits to
perform specific calculations. The first 
step is to evaluate the operators $T_n$  for the model under study, here
dimerized spin chains.
Then we calculate the ground state energy, the energy gap  and the one-magnon 
dispersion of the one-dimensional Heisenberg antiferromagnetic $S=\frac{1}{2}$
 chain. This is again done by implementing the equations on computer.

\subsection{General Equations}
The explicit form of the operators $T_n$ has to be determined so that
\begin{equation}
\label{HS_Tn}
  H_{\rm S}=T_{-2}+T_{-1}+T_0+T_1+T_2
\end{equation}
(cf.~Eqs.~(\ref{outset1},\ref{HS_1}). 
Let us consider one addend of $H_{\rm S}$ as  starting point
\begin{equation}
\label{HS_summand}
  {\bf S}_{2i}{\bf S}_{2i-1}+\alpha\left({\bf S}_{2i}{\bf S}_{2i-2}+
{\bf S}_{2i-1}{\bf S}_{2i+1}\right) \ .
\end{equation}
Obviously, only neighbouring dimers are affected. For simplicity we first 
calculate the matrix elements
\begin{equation}
\label{matrix_el}
  \langle x_{i-1},x_i|{\bf S}_{2i}{\bf S}_{2i-1}|x_{i-1},x_i \rangle
\end{equation}
where $x_{i-1},x_i\in\{s,t^1,t^0,t^{-1}\}$ are singlets or one of the triplets
 on the adjacent dimers $i-1$ and $i$. The superscript $n\in\{0,\pm1\}$
 in $t^n$ stands for the $S^z$ component. For some fixed value of $i$ we write 
\begin{equation}
\label{ohne_frust}
 {\bf S}_{2i}{\bf S}_{2i-1} = {\mathcal{T}}_{-2}+{\mathcal{T}}_{-1}+
{\mathcal{T}}_{0}+{\mathcal{T}}_{1}+{\mathcal{T}}_{2}
\end{equation}
requiring that the subscript indicates the net change of the number of
triplets.  In other words
matrix elements connecting a ket of two singlets with a bra of 
two triplets belong to ${\mathcal{T}}_{ 2}$ and those connecting a ket of one 
singlet and one triplet with a bra of two triplets belong to 
${\mathcal{T}}_{ 1}$ and so on.
 In this way one finds all the 
${\mathcal{T}}_{ n}$ and their sum is ${\bf S}_{2i}{\bf S}_{2i-1}$. 
Table \ref{matrix_tab} summarizes the results. For later convenience we split
 ${\mathcal T}_0 = {\mathcal T}_0^a+{\mathcal T}_0^b$.
All other matrix elements can be constructed by using the relation 
${\mathcal{T}}_{ n}^{\dagger}={\mathcal{T}}_{-n}$.
To incorporate the effect of frustration
 it is sufficient to note that a triplet
 is invariant under spin exchange whereas a singlet acquires a factor -1. Let
\begin{equation}
\label{mit_frust}
  \alpha({\bf S}_{2i}{\bf S}_{2i-2}+{\bf S}_{2i-1}{\bf S}_{2i+1}) = 
\sum_{n=-2}^2{\mathcal{T}}'_{n}\ .
\end{equation}
\begin{table}[bht]{}
\begin{tabular}{|ccc|}
&$4{\mathcal{T}}_{0}^a$&\\
\hline\hline
$|t^{0,\pm1},s\rangle$ & $\longrightarrow$ & 
$-|s,t^{0,\pm1}\rangle$\\
\hline
\hline
&$4{\mathcal{T}}_{0}^b$&\\
\hline\hline
$|t^0,t^{\pm1}\rangle$ & $\longrightarrow$ & 
$|t^{\pm1},t^0\rangle$\\
$|t^{\pm1},t^{\pm1}\rangle$ & $\longrightarrow$ & 
$|t^{\pm1},t^{\pm1}\rangle$\\
$|t^{\pm1},t^{\mp1}\rangle$ & $\longrightarrow$ & 
$|t^0,t^0\rangle-|t^{\pm1},t^{\mp1}\rangle$\\
$|t^0,t^0\rangle$ & $\longrightarrow$ & 
$|t^1,t^{-1}\rangle+|t^{-1},t^1\rangle$\\
\hline
\hline
&$4{\mathcal{T}}_{ 1}$&\\
\hline\hline
$|s,t^{1}\rangle, |t^{1},s\rangle$ & $\longrightarrow$ & 
$|t^{1},t^0\rangle-|t^{0},t^1\rangle$\\
$|s,t^{0}\rangle, |t^{0},s\rangle$ & $\longrightarrow$ & 
$|t^{1},t^{-1}\rangle-|t^{-1},t^1\rangle$\\
$|s,t^{-1}\rangle, |t^{-1},s\rangle$ & $\longrightarrow$ & 
$|t^{0},t^{-1}\rangle-|t^{-1},t^0\rangle$\\
\hline
\hline
&$4{\mathcal{T}}_{ 2}$&\\
\hline\hline
$|s,s\rangle$ & $\longrightarrow$ & 
$|t^{1},t^{-1}\rangle-|t^0,t^0\rangle+|t^{-1},t^1\rangle$
\end{tabular}
\caption{\label{matrix_tab} Action of the operators 
${\mathcal{T}}_{i}$ as defined by Eq.~(\ref{ohne_frust})}
\end{table}

By spin exchange the ${\mathcal{T}}'_{n}$ operators defined in
Eq.~(\ref{mit_frust}) are reduced to the 
${\mathcal{T}}_{i}$ as given in Table \ref{matrix_tab}
\begin{mathletters}
  \label{Tau_prime}
  \begin{eqnarray}
    {\mathcal{T}}'_{\pm2}&=&-2\alpha\cdot{\mathcal{T}}_{\pm2}\\
    {\mathcal{T}}'_{\pm1}&=&0\\
    {\mathcal{T}}'^a_{0}&=&-2\alpha\cdot{\mathcal{T}}_{0}^a\\
    {\mathcal{T}}'^b_{0}&=&2\alpha\cdot{\mathcal{T}}_{0}^b \ .
  \end{eqnarray}
\end{mathletters}
Finally, Eq.~(\ref{HS_Tn}) implies 
\begin{mathletters}
  \label{T_def}
  \begin{eqnarray}
    T_{\pm2} &=& \sum_{i=0}^{\frac{L}{2}-1}(1-2\alpha)\cdot{\mathcal{T}}_{\pm2}
 \\
    T_{\pm1} &=& \sum_{i=0}^{\frac{L}{2}-1}{\mathcal{T}}_{\pm1}\\
    T_{0} &=& \sum_{i=0}^{\frac{L}{2}-1}(1-2\alpha)\cdot{\mathcal{T}}^a_{0}+(1+
2\alpha)\cdot{\mathcal{T}}^b_{0} \ .
  \end{eqnarray}
\end{mathletters}
The subsequent subsection shows
 how the $T_{n}$ operators can be implemented. 
\pagebreak

For 
the antiferromagnetic $S=\frac{1}{2}$ Heisenberg chain given by 
Hamiltonian (\ref{H_ges}) we can then calculate the ground state energy by
\begin{equation}
\label{ground_en}
  E_0=\langle 0|H_{\rm{eff}}|0\rangle
\end{equation}
where we used the shorthand $|0\rangle:=\left|\prod_{i}s_i\right\rangle$
for the product state of singlets on all dimers which can be viewed
as triplet vacuum.

The actual calculations are done on finite clusters. From the linked
cluster theorem we know that the finite order contribution of
 a short-ranged perturbation does not depend on the cluster size
for sufficiently large clusters. In our one dimensional example
only neighbouring dimers are linked.  So it is sufficient
to consider 11 dimers to avoid a wrap-around in order 10, i.e.
for 11 dimers or more we are sure to find the thermodynamic
contribution. Moreover, we can check the size-independence
explicitly.

To calculate the  one magnon dispersion we have to consider the
subspace with exactly one single singlet being excited to a triplet.
Using
$|j\rangle=|s,s,\ldots,t,\ldots,s\rangle$ with one triplet $t$ ($S^z$
component does not matter) on  dimer $j$ we compute the action
of effective Hamiltonian (\ref{effhamilfin}) on $|j\rangle$. Since 
the number of triplets is conserved by construction the initial
triplet can only be shifted.
\begin{equation}
  \label{hopp_coef}
  H_{\rm{eff}}|j\rangle=J\cdot\sum_i a_i|j+i\rangle \ .
\end{equation}
The fact that the coefficients $a_i$ do not depend on $j$ relies
on the translational invariance. All dimers are equal. But this is
{\em not} necessary for our perturbation scheme to hold.
On the contrary, we consider it one of the major advantages of
the scheme presented here that it can be done in real space
without knowing the form of the eigen states in advance.

In 10$^{\mbox{th}}$ order the variable $i$ in Eq.~(\ref{hopp_coef}) runs from
-10 to 10. If $j$ is chosen appropriately, a chain segment with 11 dimers
suffices to compute all the coefficients $a_i$. For instance, for 
$a_{10}$ one has to take $j=0$ whereas for $a_0$ the right choice
is $j=5$. All sites that can be reached within 10 hops starting at
$j$ and ending at $j+i$ must be contained in the cluster to avoid
finite size effects.

Of course, given translational invariance we know that spatial
Fourier transform provides the eigen states $|k\rangle=\sqrt{2/L}\sum_i
\exp(ikj)|j\rangle$ characterized
by their lattice momentum $k$. The corresponding eigen energies read
\begin{mathletters}
\label{disp_1}
\begin{eqnarray}
  \omega(k)&=&\left<k\right|H_{\rm{eff}}|k\rangle-E_0\\\label{disp_2}
 &=&J\cdot a_0-E_0+J\cdot \sum_{j=1}^{\infty} 2a_j \cdot \cos(jk) \ .
\end{eqnarray}
\end{mathletters}

For dimerized chains the dispersion minimum, the triplet gap, is
found at $k=0$
\begin{equation}
  \label{gap_eq}
  \Delta = \omega(k=0) \ .
\end{equation}

It is beyond the scope of the present paper to compute the effective
Hamiltonian in the two-triplet sector. But an outlook on this issue
is in order. Conventional perturbation schemes have difficulties
to compute properties for two elementary particles because the structure
of the eigen states is not known beforehand. In particular, bound states of two
elementary particles are extremely difficult to obtain by conventional
perturbative methods. But if the action of the effective Hamiltonian (\ref{effhamilfin})
on two triplets is calculated everything else can be deduced.
First, one has to determine the coefficients $A_{i_1,i_2;j_1,j_2}$
defined by
\begin{equation}
H_{\rm eff} |j_1,j_2\rangle = 
\sum_{i_i,i_2} A_{i_1,i_2;j_1,j_2}|i_1,i_2\rangle\ ,
\end{equation}
where we assume that $|j_1,j_2\rangle$ has a triplet $t^1$ on dimer $j_1$
and a  triplet $t^{-1}$ on dimer $j_2$. This is sufficient to compute
triplets coupled to $S_{\rm tot}=1$  as was done successfully
in Ref. \cite{uhrig98c}; the wave
function $\psi(j_1,j_2)$ is antisymmetric under exchange 
$j_1\leftrightarrow j_2$. Second, one has to diagonalize
the matrix defined by the coefficients $A_{i_1,i_2;j_1,j_2}$. 
This can be carried out by standard Lanzcos algorithms.

Spectral functions are also accessible by flow equation perturbation
if the observable $Q$ under study is unitarily transformed by the same
transformation as the Hamiltonian
\begin{equation}
\label{flow-observ}
\frac{dQ(\ell)}{d\ell} = [\eta(\ell), Q(\ell)]\ .
\end{equation}
This is needed to know the 
matrix elements after the transformation. Preliminary studies showed that
the treatment of Eq.~(\ref{flow-observ}) is feasible within the
perturbative approach. The necessary programmes are very similar to those
for the Hamiltonian. Yet the treatment of observables
is more laborious than the one of the Hamiltonian itself.

\subsection{Computer Aided Evaluation}
\begin{figure}[hb]
\begin{picture}(5.2,2.4)
\put(0,-0.2){\includegraphics[width=8.4cm]{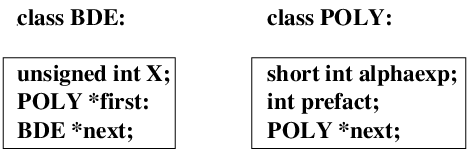}}
\end{picture}
\vspace*{3mm}
\caption{Sketch of the basic data element class}
\label{classes_fig}
\end{figure}

Again C++ is the programming language of our choice.
To encode the states of the dimerized chain it suffices to reserve two bits 
per dimer. For instance, four bytes can encode the state of a chain 
with 32 sites. The lowest bit  represents site
1, the second lowest represents site 2 and so on. By applying the $T_i$
in Eqs.~(\ref{T_def})
 these states acquire polynomials in $\alpha$ as prefactors. Thus we choose 
the basic data elements to be objects of the class BDE as sketched in 
Fig.~\ref{classes_fig}. 

The innermost part of the algorithm is the implementation of the 
${\mathcal{T}}_i$ matrices. Each matrix element is represented by a block of 
C++ code. These blocks allocate an appropriate number of basic data elements
 initialized by the current state of the chain. Then they
 modify those bits which 
represent the pair of adjacent dimers under study according to the rules of 
Table \ref{matrix_tab}. The $T_i$ are implemented in five steps:

\noindent
(i) The number of sites $L$ is chosen  even. We introduce the loop variable
\begin{equation}
  \label{p_menge}
  p \in \{1,2,\ldots,\frac{L}{2}\}
\end{equation}
such that $p=1$ addresses dimer 0 together with dimer 1, $p=2$ addresses dimer
 2 and dimer 1 and so on. Periodic boundary conditions  are used, i.e.
$p=L/2$ addresses dimer $L/2-1$ with dimer 0.\\
(ii) From $p$ a four byte long bit mask is constructed such that the bits
refering to the dimers addressed by $p$ are set to unity and zero otherwise.
 Thus $p=1$ yields $(1,1,1,1,0,\ldots,0)$ and $p=2$
 yields $(0,0,1,1,1,1,0,\ldots,0)$ and so on .\\
(iii) The appropriate part $x$ representing the two adjacent dimers is cut out
 by applying the logical AND operation: (bit mask) AND ($X$ representing the 
chain state).\\
(iv) The decimal value of $x \in \{0,1,\ldots,15\}$ is used to jump
in a SWITCH($x$)-CASE sequence to the appropriate CASE-block, 
encoding the matrix element of  ${\mathcal{T}}_i$.\\
(v) The appropriate prefactor is multiplied
 to the allocated and modified data elements.\\
Finally  $p$ is incremented by one and one goes back
 to (i) as long as $p$ meets  condition (\ref{p_menge}). 

Following these steps one obtains a sum of basic data elements
describing an elementary  chain state after the application of a $T_i$.
 In general, the complete state after the application of a $T_i$ is a
 linear combination of such elementary states encoded by the unsigned 
integers $X$. The linear combination is stored as a list.
 Since this list after its generation 
may contain identical states it is sorted according to
increasing values of $X$ by a quick-sort algorithm. Prefactors of identical 
states are added so that memory usage is reduced.

To calculate products of $T_i$, the algorithm described above
is applied repeatedly. If the chain state is a linear combination the
algorithm is applied to each addend of this linear combination.

To calculate the ground state energy (\ref{ground_en}) the expression 
$\lambda^k\sum_{|\underline{m}|=k} C(\underline{m})T(\underline{m})$ from
Eq.~(\ref{effhamilfin}) must be applied to the triplet vacuum $|0\rangle$
 for each order k.
First the current index $\underline{m}$ and its coefficient $C(\underline{m})$
are read from an input file prepared previously, see Sect. II.B.
Then $T(\underline{m})$  is applied to $|0\rangle$ as described above.
 Since $M(\underline{m})=0$
$T(\underline{m})$ reproduces $|0\rangle$ up to a prefactor
 which is a polynomial in $\alpha$. This polynomial and $C(\underline{m}$)
are piped  to an algebraic computer programme (Maple) 
to multiply them. This scheme 
is iterated with intermediate summations by Maple 
till the final result is found.

The result for the ground state energy per spin of the Hamiltonian
 (\ref{H_ges}) up to 10$^{\mbox{th}}$
 order in $\lambda$ reads
\begin{eqnarray}
\nonumber
\lefteqn{\frac{\epsilon_0}{J} = \left(1 - 2{}\alpha \right)^2
\bigg(
- {\displaystyle \frac {3}{4}} {}\overline{\lambda} ^{2}
- \bigg( {\displaystyle \frac {3}{4}}  
+ {\displaystyle \frac {3}{2}} {}\alpha \bigg)
{}\overline{\lambda} ^{3}  
- \bigg({\displaystyle \frac {13}{16}}
+ {\displaystyle \frac {27}{4}} {}\alpha}
\\ \nonumber
& &
- {\displaystyle \frac {3}{4}} {}\alpha ^{2}\bigg){}\overline{\lambda} ^{4} 
- \bigg(
 {\displaystyle \frac {89}{48}}  
+ {\displaystyle \frac {311}{24}} {}\alpha  
+ {\displaystyle \frac {93}{4}} {}\alpha ^{2} 
- {\displaystyle \frac {45}{2}} {}\alpha ^{3}\bigg){}\overline{\lambda} ^{5}
- \bigg( {\displaystyle \frac {463}{96}}
\\ \nonumber
& &
+ {\displaystyle \frac {227}{9}} {}\alpha 
+ {\displaystyle \frac {1307}{12}} {}\alpha ^{2} 
- 42{}\alpha ^{3} 
- {\displaystyle \frac {159}{2}} {}\alpha ^{4}\bigg){}\overline{\lambda}^{6}
- \bigg({\displaystyle \frac {81557}{6912}}
\\ \nonumber
& & 
+ {\displaystyle \frac {257909}{3456}} {}\alpha  
+ {\displaystyle \frac {215995}{864}} {}\alpha ^{2}
+ {\displaystyle \frac {173579}{432}} {}\alpha ^{3} 
- {\displaystyle \frac {14865}{16}} {}\alpha ^{4}
\\ \nonumber
& & 
+ {\displaystyle \frac {879}{8}} {}\alpha ^{5}  
\bigg)
\overline{\lambda} ^{7}
- \bigg( {\displaystyle \frac {414359}{12960}}
+ {\displaystyle \frac {139801}{648}} {}\alpha 
+ {\displaystyle \frac {8477587}{12960}} {}\alpha ^{2} 
\\ \nonumber
& & 
+ {\displaystyle \frac {152558}{81}} {}\alpha ^{3} 
- {\displaystyle \frac {2774357}{1620}} {}\alpha ^{4}
- 4002{}\alpha ^{5} 
+ {\displaystyle \frac {4527}{2}} {}\alpha ^{6}\bigg)
\overline{\lambda} ^{8}
\\ \nonumber
& & 
- \bigg( {\displaystyle \frac {2354594813}{24883200}}  
+ {\displaystyle \frac {7341879263}{12441600}} {}\alpha 
+ {\displaystyle \frac {14053262981}{6220800}} {}\alpha ^{2} 
\\ \nonumber
& & 
+ {\displaystyle \frac {1591335559}{345600}} {}\alpha ^{3}
+ {\displaystyle \frac {9560574943}{1555200}} {}\alpha ^{4}
- {\displaystyle \frac {8121212969}{259200}} {} \alpha ^{5}
\\\nonumber
& &  
+ {\displaystyle \frac {453741}{64}} {}\alpha ^{6}
+ {\displaystyle \frac {248391}{32}} {} \alpha ^{7} 
\bigg)\overline{\lambda} ^{9} 
- \bigg({\displaystyle \frac {106469295871}{373248000}} 
\\\nonumber 
& & 
+ {\displaystyle \frac {82849717337}{46656000}} {} \alpha
+ {\displaystyle \frac {107584683283}{15552000}} {}\alpha ^{2}
+ {\displaystyle \frac {89796462557}{5832000}} {}\alpha ^{3}
\\\nonumber
& & 
+ {\displaystyle \frac {160938279937}{5832000}} {}\alpha ^{4}
- {\displaystyle \frac {57686123141}{972000}} {}\alpha ^{5}
\\\label{E_grund}
& &   
- {\displaystyle \frac {143920286959}{972000}} {}\alpha ^{6}\!
+ {\displaystyle \frac {339171}{2}} {}\alpha ^{7}\!
- {\displaystyle \frac {336527}{16}} {}\alpha ^{8} 
\bigg)\overline{\lambda} ^{10} \!
\bigg) ,
\end{eqnarray}
where the shorthand $\overline{\lambda} = \frac{1}{4}\lambda$ is used.

To calculate the dispersion $\omega(k)$  the hopping
 elements $a_i$ in Eq.~(\ref{disp_2}) have to be determined. 
The effect of $T(\underline{m})$ with
 $M(\underline{m})=0$ is to shift a triplet by at most $|\underline{m}|$
 dimers. Hence it suffices to perform the calculation for a given $a_i$ on an 
appropriate chain segment. The calculations are analogous to those
 for the ground state energy. Note that by Eq.~(\ref{hopp_coef})
 the general effect of a product $T(\underline{m})$ will be a sum of
 states each containing one triplet on different dimers. For a given $a_i$
 the corresponding state has to be found in that list. Its
polynomial prefactor yields $a_i$. Results for
 the hopping elements up to 10$^{\mbox{th}}$ order in $\lambda$
 are presented in appendix \ref{app-effhopp}. 

The calculations took 10h-30h strongly depending on the number of
 sites $L$; 100MB-500MB memory were used. The computations were done on a Sun
 Ultra workstation.

\section{Results and Comparison to Other Methods}
So far all results refer to Hamiltonian (\ref{hamil2}). In this section we 
prefer to present the results corresponding to Hamiltonian (\ref{hamil1}) by
 substituting according to Eq.~(\ref{substitution}). Since 
$\lambda$ is the expansion parameter it is substituted {\em before} we 
manipulate the equations further, see below. The frustration $\alpha$ on the 
other hand is treated as a fixed parameter throughout further manipulations. It
 is substituted only at the very end. 
\subsection{Ground State Energy}
\begin{figure}
\begin{picture}(8.2,5.4)
\put(0,-0.2){\includegraphics[width=8.4cm]{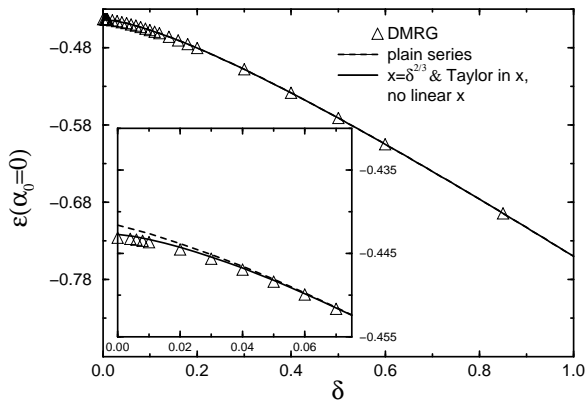}}
\end{picture}
\caption{Ground state energy per site $\epsilon$ vs. dimerization $\delta$
 without frustration ($\alpha_0=0$) for Hamiltonian (\ref{hamil2}). For details
 see main text. The inset shows an enlargement for small $\delta$.}
\label{E0_delta_1_fig}
\end{figure}

Substituting $J$, $\lambda$ and $\alpha$ in Eq.~(\ref{E_grund}) leads to the
 dashed curve in Fig.~\ref{E0_delta_1_fig} called ``plain series''. The solid
 line represents a more sophisticated approach. Substituting $J$ and $\lambda$
 leads to an expression in $\delta$ and $\alpha$. 
In order to profit from the knowledge that the ground state energy is lowered
by external dimerization as $\Delta E_0 \propto - \delta^{4/3}$ \cite{cross79}
$\delta=x^{\frac{3}{2}}$ is substituted. Thereby $\alpha$ is treated as
fixed parameter which does not change the exponents, here $4/3$. 
Their values are protected by symmetry \cite{uhrig96b}
as long as logarithmic corrections are neglected \cite{black81,affle89}.
 This means that this
statement is true as long as $\alpha$ is below its critical
value $\alpha_c=0.241167(\pm 5)$ \cite{egger96}. Above this value
the translational invariance is broken spontaneously and the ground
state is dimerized so that the leading power becomes linear
$\Delta E_0 \propto - \delta$ (see below).

The result for the ground state energy is expanded in
 a Taylor series in $x$ up to order 10 about $x=1$. 
This is the limit of strong dimerization for which our 
expansion holds. Adding the term $Y\cdot(1-x)^{11}$ 
introduces an unknown $Y$ which can be fitted such that the linear term in $x$
 vanishes. In this way the leading order contribution to the energy lowering
agrees with the continuum prediction. The result of this procedure 
is depicted as solid curve in Fig.~\ref{E0_delta_1_fig} and is labeled
accordingly. Both curves are compared to numeric results
 from Density Matrix Renormalization Group (DMRG) shown as symbols.
The agreement is excellent.  The accordance is better for the biased
extrapolation based on the continuum theoretical power law as was to
be expected. The extrapolated results can be trusted even quantitatively
down to 1 or 2\%.

In absence of frustration ($\alpha=\alpha_0=0$) the right hand side of
 Eq.~(\ref{E_grund}) becomes a polynomial in $\lambda$ of degree 10 with 
rational coefficients. Using a different perturbation method Barnes 
{\it et al.} \cite{barne98} calculated the same coefficients up to
 order 9, providing a good check. Furthermore,
 Gelfand {\it et al.} \cite{gelfa90} also calculated a polynomial in 
$\lambda$. Their method gives the coefficients as real numbers up to 
order 15. We could varify these numbers to the given precision in 
Ref.~\cite{gelfa90} up to order 10.

\begin{figure}
\begin{picture}(8.2,5.4)
\put(0,-0.2){\includegraphics[width=8.4cm]{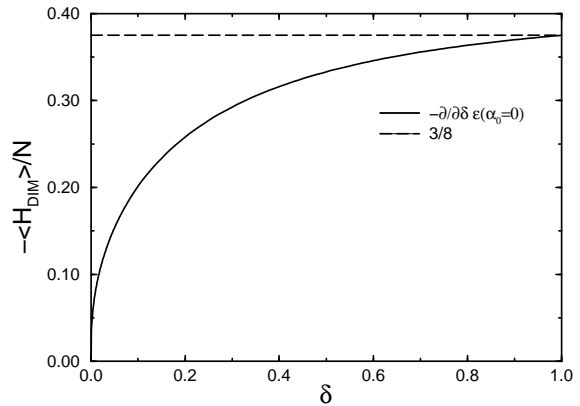}}
\end{picture}
\caption{Expectation value of the dimerization operator $H_{\rm DIM}$ vs.
 dimerization $\delta$ without frustration ($\alpha_0=0$).}
\label{<dim>_delta_1_fig}
\end{figure}

One of the major motivations to study dimerized spin chains comes from 
spin-Peierls systems where spin and lattice degrees of freedom are coupled. In
the adiabatic description of this phenomenon one adds the elastic energy
$\frac{K}{2}\delta^2$ to the Hamiltonian (\ref{hamil1}). This term takes into
account that it costs energy  to modulate the magnetic couplings.
 To determine the equilibrium value of $\delta$ the ground
state energy is minimized by variation of $\delta$
\begin{equation}
0 = \frac{\partial \epsilon_0}{\partial\delta}+K \delta \ .
\end{equation}
So one has to know the derivative of $\epsilon_0$ which is given by the
expectation value per site of the dimerization operator
$H_{\rm DIM}=(H(\delta)-H(0))/\delta$ with $H$ from Eq.~(\ref{hamil1})
\begin{equation}
  \label{H_dim}
  \frac{\langle H_{\rm DIM} \rangle}{L}=
\frac{\partial \epsilon_0}{\partial \delta} \ .
\end{equation}
In Fig.~\ref{<dim>_delta_1_fig}
we plot $\partial \epsilon_0/\partial\delta$ as derived from our results
for the ground state energy. Note that $3/8$ is an upper bound for
the expectation value of the dimerization since the dimerization is
maximum if every second bond is occupied by a singlet 
$\langle {\bf S}_{2j}{\bf S}_{2j+1}\rangle=-3/4$.
 This upper bound is excellently
complied with by our results since we expand around the limit of 
complete dimerization $\delta=1$.

\begin{figure}
\begin{picture}(8.2,11.4)
\put(0,-0.2){\includegraphics[width=8.4cm]{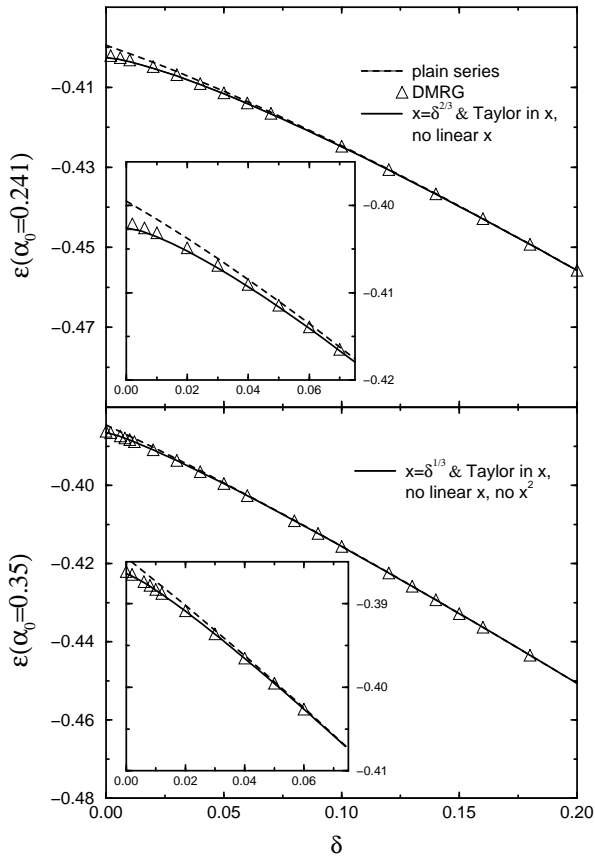}}
\end{picture}
\caption{Ground state energy $\epsilon$ vs. dimerization $\delta$;
 frustration $\alpha_0=0.241$ (upper panel), $\alpha_0=0.35$ (lower panel)}.
The insets show enlargements for small $\delta$.
\label{E0_delta_2_fig}
\end{figure}

\begin{figure}
\begin{picture}(8.2,11.8)
\put(0,-0.2){\includegraphics[width=8.4cm]{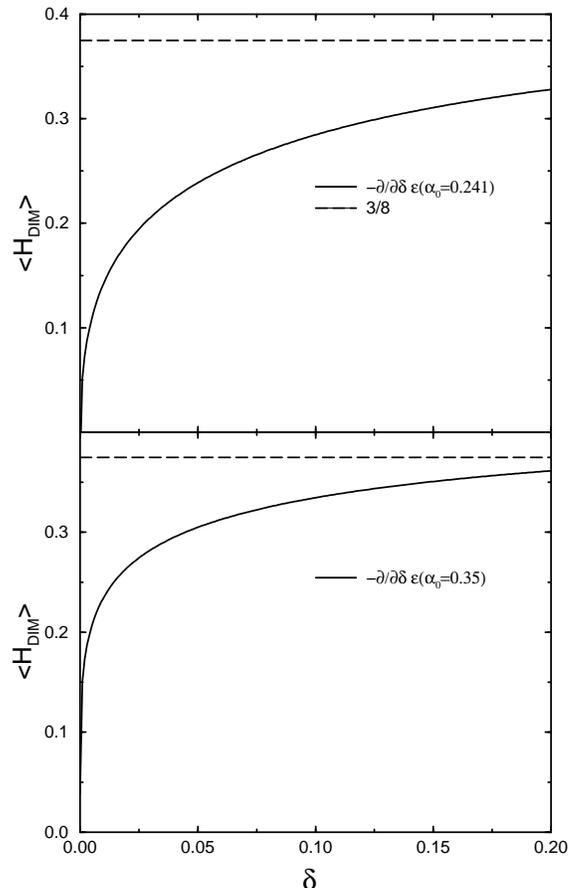}}
\end{picture}
\caption{Expectation value of the dimerization operator $H_{\rm DIM}$ vs.
 dimerization $\delta$; frustration $\alpha_0=0.241$ (upper panel),
 $\alpha_0=0.35$ (lower panel).}
\label{<dim>_delta_2_fig}
\end{figure}

Fig. \ref{E0_delta_2_fig} show the  ground state energy per site
 for $\alpha_0=0.241$ and $\alpha_0=0.35$. For supercritical
 frustration we know that the leading term is linear in $\delta$. In order 
to be able to describe well 
the crossover from $\delta^{\frac{4}{3}}$ to $\delta$
 behavior we use the substitution $\delta=x^{3}$ and suppress linear and
 quadratic terms in the  Taylor expansion by appropriate higher order 
terms in $(1-x)$. This procedure
leads to a series in $\delta$ comprising terms $\propto \delta$ and
 $\propto \delta^{\frac{4}{3}}$ as well as higher terms in $\delta^{1/3}$.
The agreement obtained in comparison to DMRG data is again excellent
down to very low values of dimerization. In Fig.~\ref{<dim>_delta_2_fig}
the expectation values of the corresponding dimerization operators
$H_{\rm DIM}$ are plotted. 

We limited the range of $\delta$ for finite frustration  to the interval 
$[0,0.2]$. This is required by the nature of our perturbative expansion
around isolated dimers. Fixing $\alpha_0$ (not $\alpha$ !) at a finite value
 implies that the limit $\delta \rightarrow 1$ is {\em not} the limit of 
isolated dimers. The point $\delta=1$ corresponds to a spin ladder with
 coupling $2J_0$ on the rungs and $\alpha_0 J_0$ on the legs.
Of course, our bare perturbative results pertain also to the ladder where
they correspond to an expansion around the rung limit. This limit has
already been investigated intensively \cite{weiho98} so that we refrain here
from a comprehensive analysis. Note that the perturbation  in the ladder
is considerably simpler since $N=1$ in Eq.~(\ref{outset1})
whereas $N=2$ is treated here. In other words there is no creation or
annihilation of {\em two} triplets but only shifts of triplets or 
creation or annihilation of one triplet.

To understand the reason for the restricted applicability of the analysis
presented here a simple comparison of couplings suffices.
Using  $x=\delta^{\frac{2}{3}}$ or $x=\delta^{\frac{1}{3}}$ is 
optimized to treat the case where the 
coupling $J_0(1-\delta)$ is the dominant perturbation. As a rule of thumb
 this is the case if $2\alpha_0 < 1-\delta$. Otherwise one has to treat the
 frustrating coupling as the dominant perturbation. We restrict our present
 analysis to the regime where the $1-\delta$ coupling is the dominant 
perturbation so that $\delta$ may not be chosen too large. For 
$\alpha_0\approx 0.35$ the dimerization $\delta$  should not exceed about
$0.3$.

\subsection{Energy Gap}
\begin{figure}
\begin{picture}(8.2,5.4)
\put(0,-0.2){\includegraphics[width=8.4cm]{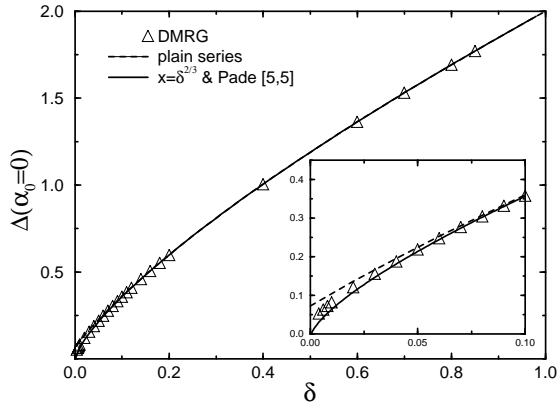}}
\end{picture}
\caption{Energy gap $\Delta$ vs. dimerization $\delta$ without frustration.
 The numbers in the square bracket stand for the
 polynomial degree in the numerator and the denominator of the Pad\'e
approximant, respectively. The inset shows an enlargement for small $\delta$.
\label{gap_delta_1_fig}}
\end{figure}

\begin{figure}
\begin{picture}(8.2,5.2)
\put(0,0){\includegraphics[width=8.4cm]{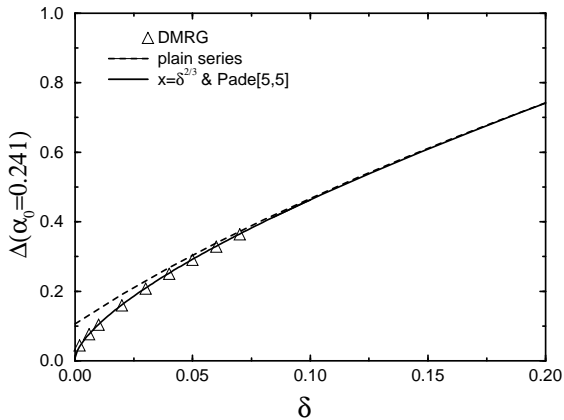}}
\end{picture}
\caption{Energy gap $\Delta$ vs. dimerization $\delta$ for critical
 frustration $\alpha_0=0.241$.}
\label{gap_delta_2_fig}
\end{figure}

\begin{figure}
\begin{picture}(8.2,11.6)
\put(0,-0.2){\includegraphics[width=8.4cm]{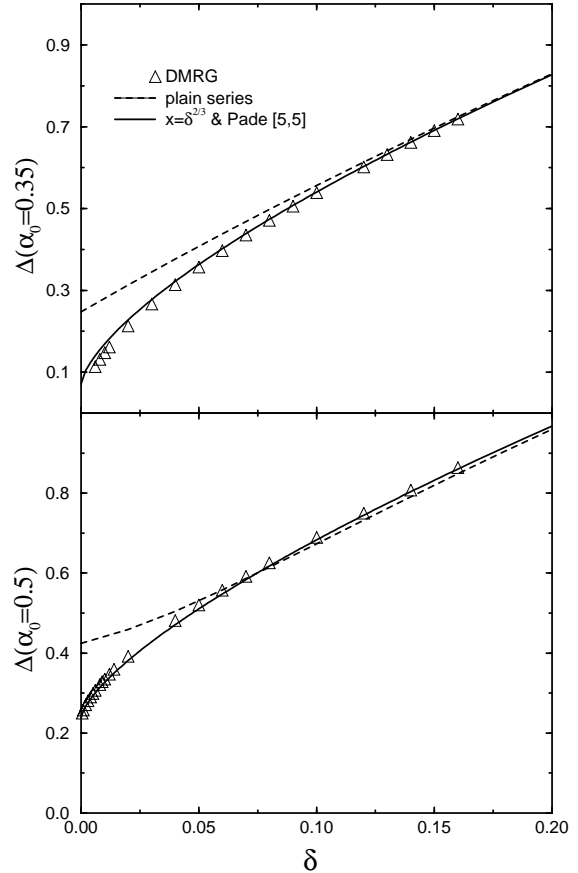}}
\end{picture}
\caption{Energy gap $\Delta$ vs. dimerization $\delta$ for supercritical
 frustration $\alpha_0=0.35$, $\alpha_0=0.5$.}
\label{gap_delta_3_fig}
\end{figure}

The first approach is again to substitute $J$, $\lambda$ and $\alpha$ in 
Eq.~(\ref{gap_eq}) according to Eq.~(\ref{substitution}). 
In the case $\alpha_0=0$ our result agrees to the $9^{\mbox{th}}$ degree
 polynomial Barnes {\it et al.} \cite{barne98} computed for the energy gap.
For various values of $\alpha_0$ the results
 are plotted as dashed curves labeled ``plain series'' in 
Figs.~\ref{gap_delta_1_fig}, \ref{gap_delta_2_fig} and \ref{gap_delta_3_fig}.

\begin{figure}
\begin{picture}(8.2,5.2)
\put(0,-0.1){\includegraphics[width=8.4cm]{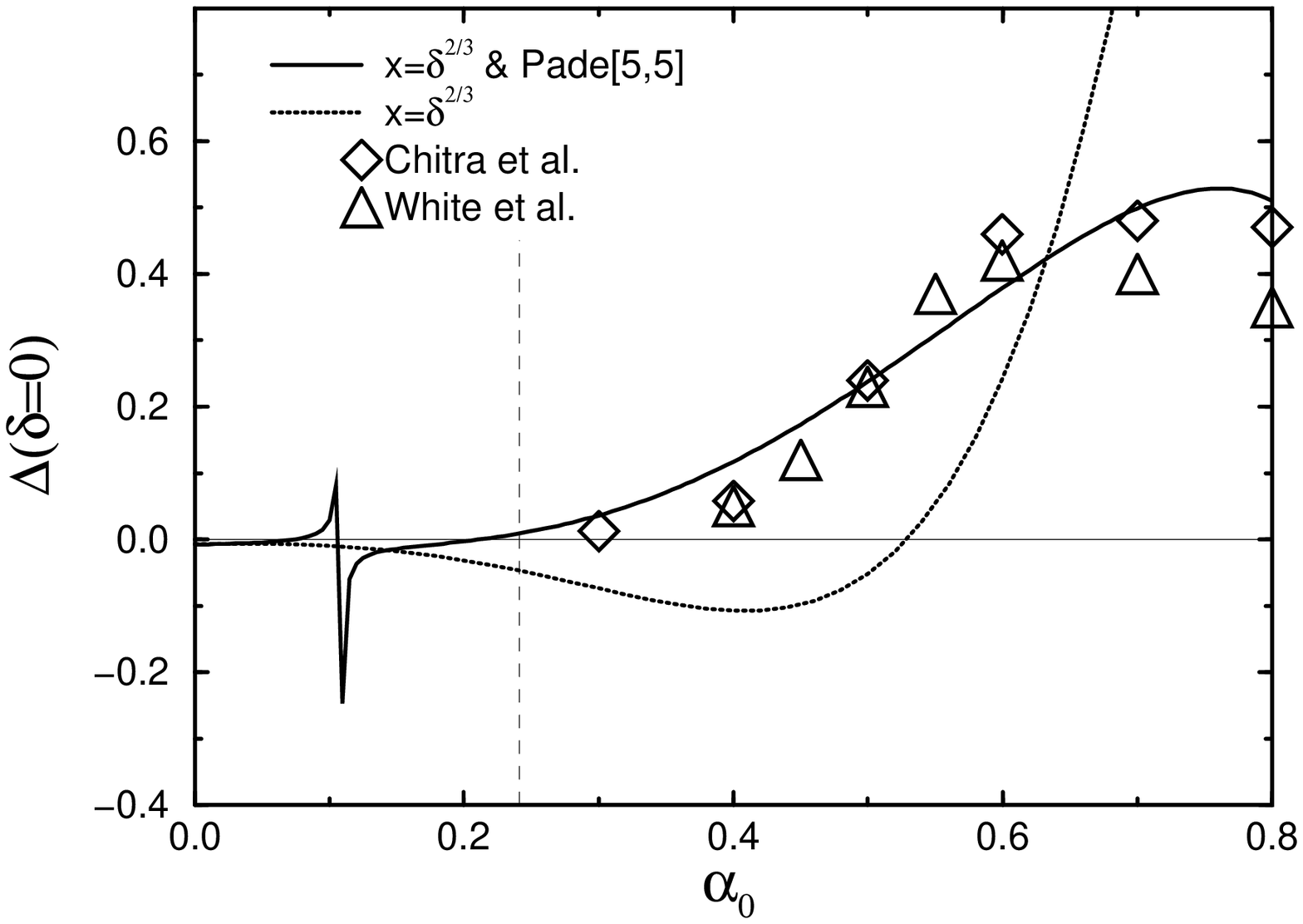}}
\end{picture}
\caption{Energy gap $\Delta$ vs. frustration $\alpha$ at $\delta=0$. 
The dashed vertical line indicates the value of $\alpha_c$.}
\label{gap_alpha_fig}
\end{figure}

The solid curves were obtained by an extrapolation biased by the
continuum theory results $\Delta \propto \delta^{2/3}$ 
for subcritical frustration \cite{cross79} and 
$\Delta -\Delta|_{\delta=0} \propto \delta^{2/3}$ for supercritical
frustration, see e.g.~Ref.~\cite{uhrig99a}.
To this end, we substituted $J$ and $\lambda$ in 
Eq.~(\ref{gap_eq}) according to Eq.~(\ref{substitution}).
 Next  $\delta=x^{\frac{3}{2}}$ is replaced treating 
$\alpha$ as a fixed parameter. This result is  re-expressed by
an appropriate Pad\'e approximant about $x=1$, which 
corresponds to the dimer limit.
As will be seen below the use of a Pad\'e approximant instead
of a series in $x$ matters only for sizable frustration.
 The Pad\'e approximant is chosen
such that it takes the information about the 11 coefficients of
the expansion into account. In the final expression
$x=\delta^{\frac{3}{2}}$ and $\alpha=\frac{\alpha_0}{1-\delta}$
are inserted to obtain the data shown.
The comparison to  DMRG data \cite{uhrig99a} shows that the biased
extrapolation is extremely precise for  most choices of parameters.
In order to check the reliability of our results in the most difficult case
we extrapolate our results to zero dimerization. This means we employ
the above procedure of a biased Pad\'e approximant and set $\delta=0$. 
The gap dependence on the frustration $\alpha_0$ is depicted
 in Fig.~\ref{gap_alpha_fig}
and compared to DMRG results \cite{chitr95,white96}. 
For comparison we include also an extrapolation of the series in 
$x=\delta^{2/3}$.
Rigorously, there is no gap below the critical value
$\alpha_c$ \cite{halda82b}.  The numerical
value ($\alpha_c = 0.241167(\pm5)$) is taken from  Ref.~\cite{egger96}.
The wiggle at about $\delta=0.1$ is a spurious 
pole resulting from the Pad\'e approximation which is not
present in the series representation. The Pad\'e approximant, however,
is better for larger frustration.
The overall agreement is good but not excellent. Obviously, neither
the Pad\'e approximant nor the series in $x$
are fit to describe the essential
singularity at critical frustration. Yet, 
more sophisticated approximants which allow
to extrapolated in several variables (here: $\lambda$ and 
$\alpha_0$) might render a more efficient analysis of the
available expansion coefficients.

To complete the analysis of our expansion coefficients other approximants
were also used. The common approach to detect critical behaviour is
to approximate the derived logarithm of the function under study by a
Pad\'e approximant (Dlog Pad\'e). In this representation
the position $\lambda_c$ of a singularity $(\lambda-\lambda_c)^\gamma$ is
given by a pole and its residue defines the critical exponent $\gamma$.
The results do not depend much on whether the Hamiltonian (\ref{hamil1})
or (\ref{hamil2}) is used. In fact, the results for Hamiltonian
(\ref{hamil2}) agree a bit better with the predictions $\lambda_c=1$
and $\gamma=2/3$ than those for Hamiltonian (\ref{hamil1}).
In absence of frustration we find $\lambda_c=1.002$ and $\gamma=0.74$.
The value for the position is very encouraging; the result for the
critical exponent, however, is a bit disappointing. The same observation
was made in Ref.~\cite{barne98}. No significant change in the
exponent occurs if the position of the singularity $\lambda_c=1$
is pre-set.

The reason for the difficulty to find the correct exponent is found in
the logarithmic corrections. This can be seen in two ways.
The first one is to go to the critical frustration where no logarithmic
corrections are present. This proved useful in numerical
analyses \cite{chitr95}, too. Indeed, at $\alpha_0=\alpha_c$
we find $\gamma=0.65$ which agrees with $2/3$ within 3\%.
The second  is to take the logarithmic corrections into account. 
Usually, they are computed for subcritical
frustration and given in the form \cite{kadan80,black81,affle89}
\begin{equation}
\label{asymp1}
\Delta(\delta) \propto \delta^{2/3}/|\ln(\delta/\delta_0)|^{1/2}
\end{equation}
where $\delta_0$ is some unknown constant. Hence, we apply the
Dlog Pad\'e approximation to 
\begin{equation}
f(\delta):=\sqrt{-\ln(\delta) + h(\alpha_0)}\Delta\ ,
\end{equation}
where $h(\alpha_0)$ is some constant depending on the frustration.
We look at the biased approximant $f'/f=P(\delta^2)/\delta$ 
about $\delta^2=1$ where
$P(\delta^2)$ is a polynomial of order 9 in $\delta^2$ which takes
the obvious symmetry $\delta \leftrightarrow -\delta$
 of Hamiltonian (\ref{hamil1}) into account.
To be able to expand about isolated dimers the replacement
$\alpha_0 \to \alpha_0(1-\delta^2)$ is carried out.
 The value $P(0)$ then
is a direct estimate for the critical exponent $\gamma$. Assuming that
the coefficients of $P$ decrease quickly on increasing order once the
logarithmic correction is properly taken into account we fix
the constant $h(\alpha_0)$  such that the coefficient of order 9 vanishes.
In this way we find $\gamma=0.68$ and $h=5.65$ at zero frustration.
At critical frustration the values are $0.56$ and $2.50$, respectively.

The good agreement without frustration indicates that logarithmic
corrections are indeed the reason for the difficulty to 
determine the correct exponents from the perturbative data.
The unsatisfactory agreement at critical frustration tells us that
the way $h$ is determined is not optimum. It would be very helpful
if a  prediction for the gap as function of 
$\delta$ and $\alpha-\alpha_c$ existed which comprised also the
regime of supercritical frustration.

\subsection{Dispersions}

The one magnon dispersion $\omega(k)$ for the antiferromagnetic Heisenberg
 chain is given by Eq.~(\ref{disp_2}). The hopping elements $a_i$ are
 given in appendix \ref{app-effhopp}. 
For $\alpha_0=0$ they can be compared till order 5
with those computed in  Ref.~\cite{barne98}. Full agreement is found.

To deduce the dispersion relations from  the bare coefficients
several approaches will be presented.
The direct approach is again to substitute $J$, $\lambda$ and $\alpha$ in
 Eq.~(\ref{disp_2}) according to Eq.~(\ref{substitution}). 
By construction, the resulting curves reproduce the same
 energy gaps at $k=0$ as the gap results labeled ``plain series'' in
 the preceding section. We learned, however, in the preceding section
that a biased extrapolation is very useful to approximate the gaps.
So the question arises how the good biased extrapolations can be
used for the description of dispersion relations.

\begin{figure}
\begin{picture}(8.2,5.4)
\put(0,-0.2){\includegraphics[width=8.4cm]{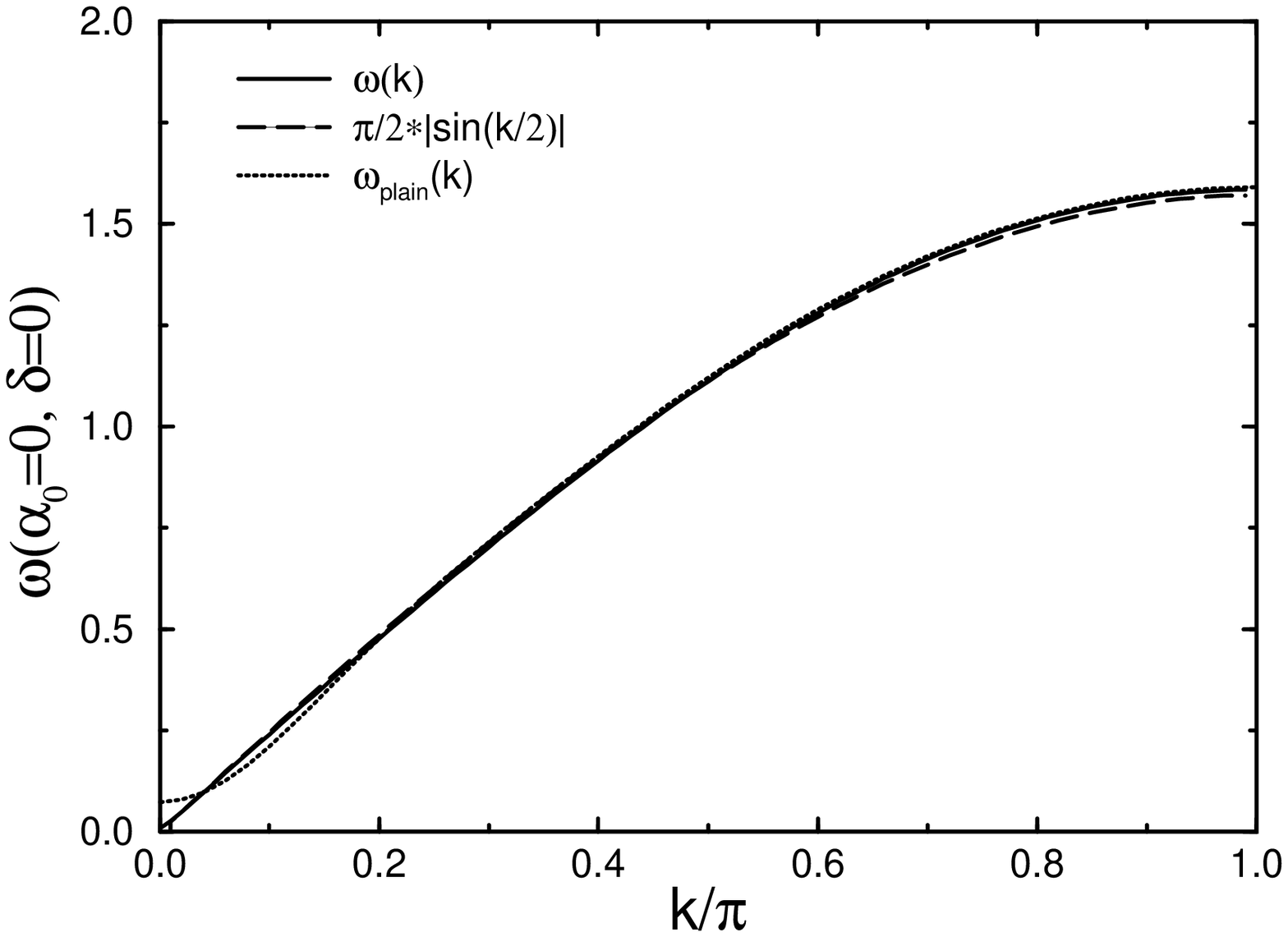}}
\end{picture}
\caption{Dispersion $\omega$ vs. wave vector $k$ without frustration and
dimerization.The curve $\omega_{\rm plain}$ depicts the plain series 
result.}
\label{omega_1_fig}
\end{figure}

\begin{figure}
\begin{picture}(8.2,5.4)
\put(0,-0.2){\includegraphics[width=8.4cm]{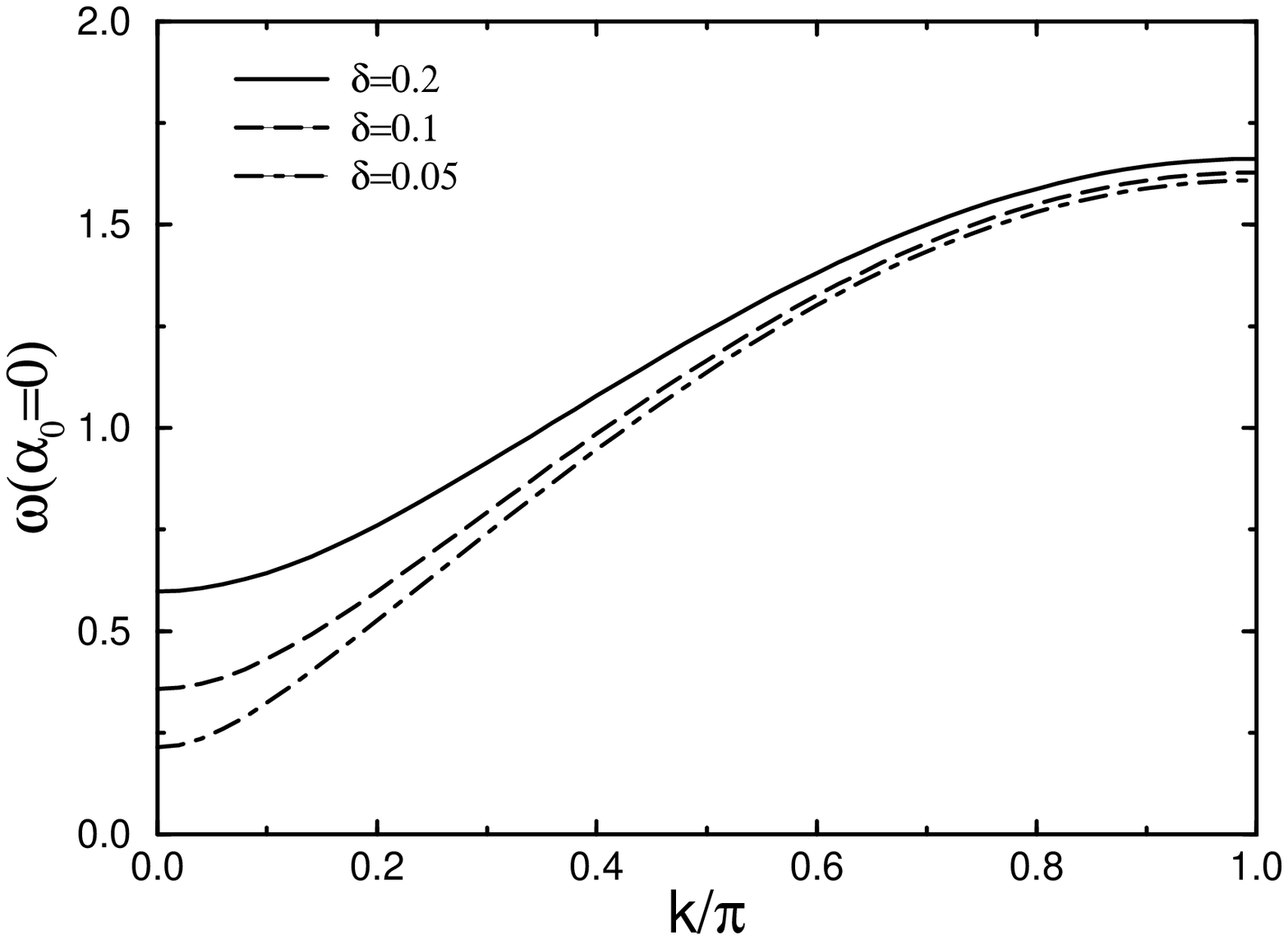}}
\end{picture}
\caption{Dispersion $\omega$ vs. wave vector $k$ at $\alpha_0=0$
and various dimerizations.}
\label{omega_2_fig}
\end{figure}

\begin{figure}
\begin{picture}(8.2,5.4)
\put(0,-0.2){\includegraphics[width=8.4cm]{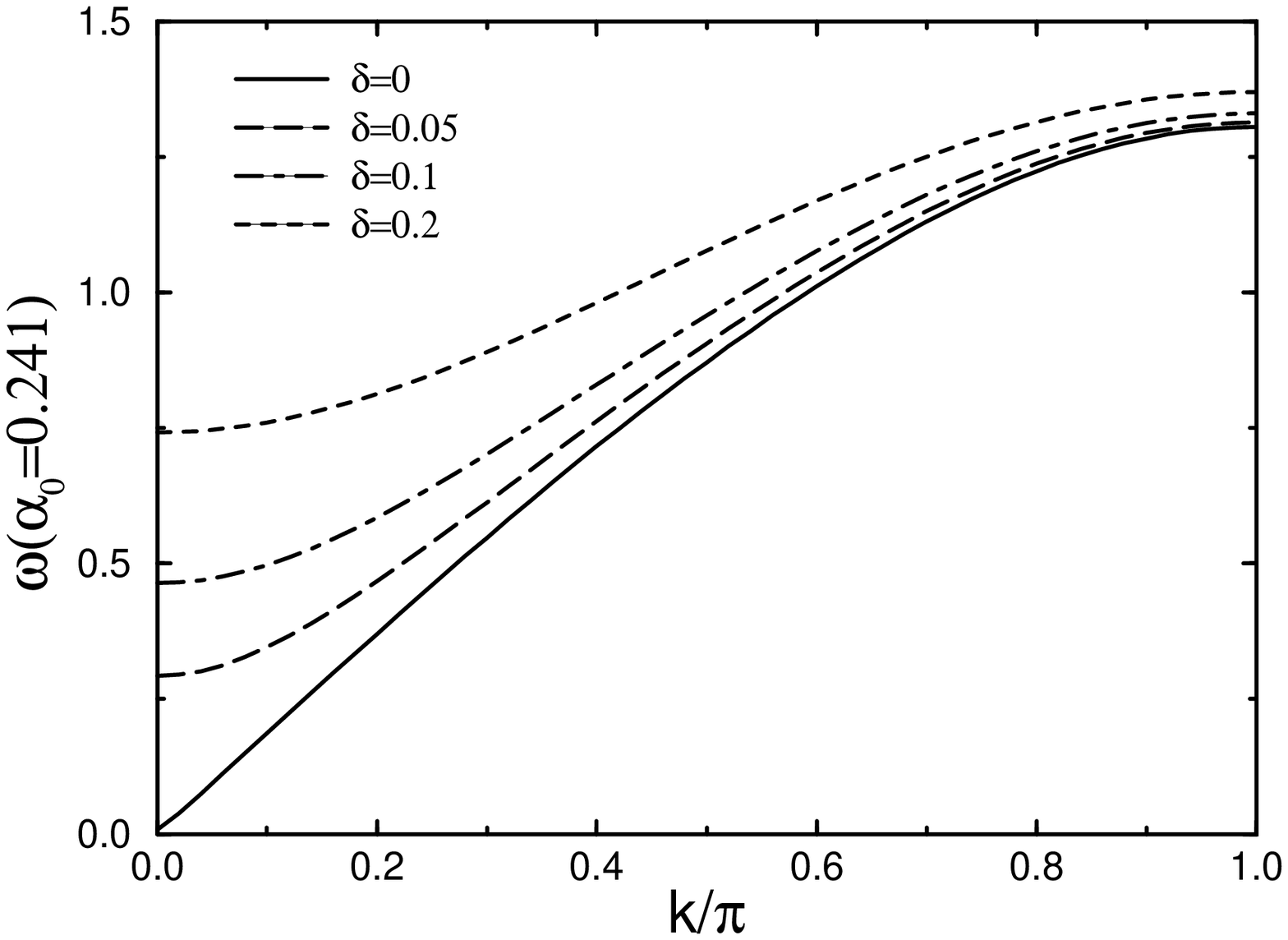}}
\end{picture}
\caption{Dispersion $\omega$ vs. wave vector $k$ at $\alpha_0=0.241$
and various dimerizations.}
\label{omega_3_fig}
\end{figure}

\begin{figure}
\begin{picture}(8.2,5.4)
\put(0,-0.2){\includegraphics[width=8.4cm]{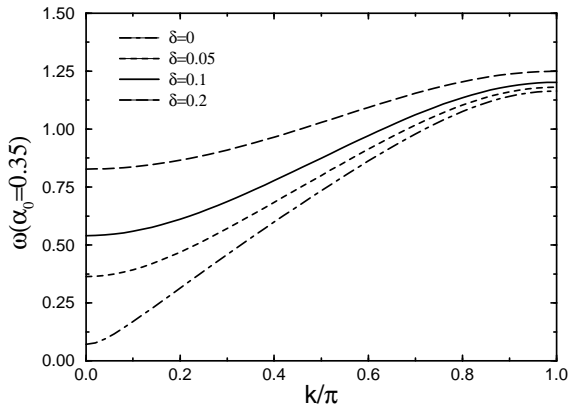}}
\end{picture}
\caption{Dispersion $\omega$ vs. wave vector $k$ at $\alpha_0=0.35$
and various dimerizations.}
\label{omega_4_fig}
\end{figure}

Motivated by the behaviour of Lorentz-invariant systems where the
dispersion passes from $\omega = v_{\rm S} k$ ($v_{\rm S}$ spin wave
velocity) to $\omega \propto
\sqrt{\Delta^2+ (v_{\rm S} k)^2}$ when a gap opens
we use the following procedure. We substitute $J$ and $\lambda$ into
 Eq.~(\ref{disp_2}) leading to  the plain series result we shall refer to as 
$\omega_{{\rm plain}}(k;\alpha)$. Then the difference   
\begin{equation}
  \label{disp_diff}
  \omega^2_{\rm diff}(k):=\omega^2_{\rm plain}(k;\alpha)-\omega^2_{\rm plain}
(k=0;\alpha)
\end{equation}
is expanded
as Taylor series about $\delta=1$ up to 10$^{\mbox{th}}$ order
 while $\alpha$ is treated as fixed parameter. 
Only then $\alpha$ is substituted according to Eq.~(\ref{substitution}).
Finally, the dispersion is computed by the quadratic mean
\begin{equation}
  \label{disp_sqrt}
  \omega(k)=\sqrt{\Delta^2(\alpha_0,\delta)+\omega^2_{\rm diff}(k)} \ .
\end{equation}
For $\delta=\alpha_0=0$ the result is shown in Fig.~(\ref{omega_1_fig})
where it is also compared to the plain series result to illustrate the
effect of the quadratic mean.
The curve is compared to the rigorous result known from Bethe ansatz
\cite{cloiz62}. 
The quadratic mean matters only for low energies.
Figs.~\ref{omega_2_fig}, \ref{omega_3_fig} and \ref{omega_4_fig} show
the corresponding results for various $\delta$ and $\alpha$ values
for relatively low frustration
in the subcritical regime or close to it.
 As soon as there is some dimerization, for
instance $\delta \ge 0.05$ in Fig.~\ref{omega_2_fig}, the difference
 between the plain series (not shown) and the quadratic mean
(\ref{disp_sqrt}) is no longer discernible.

\begin{figure}
\begin{picture}(8.2,5.4)
\put(0,-0.2){\includegraphics[width=8.4cm]{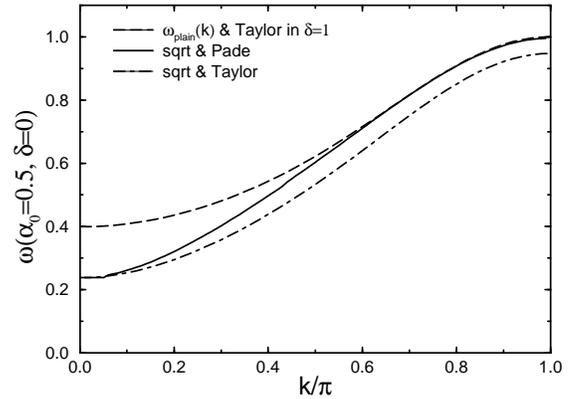}}
\end{picture}
\caption{Dispersion $\omega$ vs. wave vector $k$ at $\alpha_0=0.5$ and
 $\delta=0$. The results of
three different ways to deduce the dispersion relation from the  
bare coefficients are depicted.}
\label{omega_5_fig}
\end{figure}

Entering the regime of large frustration ($\alpha_0 > 0.4$) the calculations 
have to be modified slightly. In Fig.~\ref{omega_5_fig} we compare two 
different approaches to calculate the dispersion $\omega(k)$. The dashed 
line labeled ``$\omega_{\rm plain}(k)$ \& Taylor'' is obtained by 
expanding $\omega_{\rm plain}(k;\alpha)$ in a Taylor series about 
$\delta=1$ with  $\alpha$ as fixed parameter. Finally, $\alpha$ is
 substituted according to Eq.~(\ref{substitution}) for
$\delta=0$, $\alpha_0=0.5$. The resulting curve is far off for low
wave vectors. To get good values for the energy gap it is necessary
to use the square root representation (\ref{disp_sqrt}).
 On the other hand, the value at  $k=\pi$ is unity which is 
 the exact result \cite{caspe82,mulle99}. If the square root 
representation (\ref{disp_sqrt}) is used the dashed-dotted line
 labeled ``sqrt \& Taylor'' in Fig.~\ref{omega_5_fig} is obtained.
The energy gap is nicely reproduced \cite{mulle99} but the agreement
at the dispersion maximum deteriorates considerably.

\begin{figure}
\begin{picture}(8.2,5.4)
\put(0,-0.2){\includegraphics[width=8.4cm]{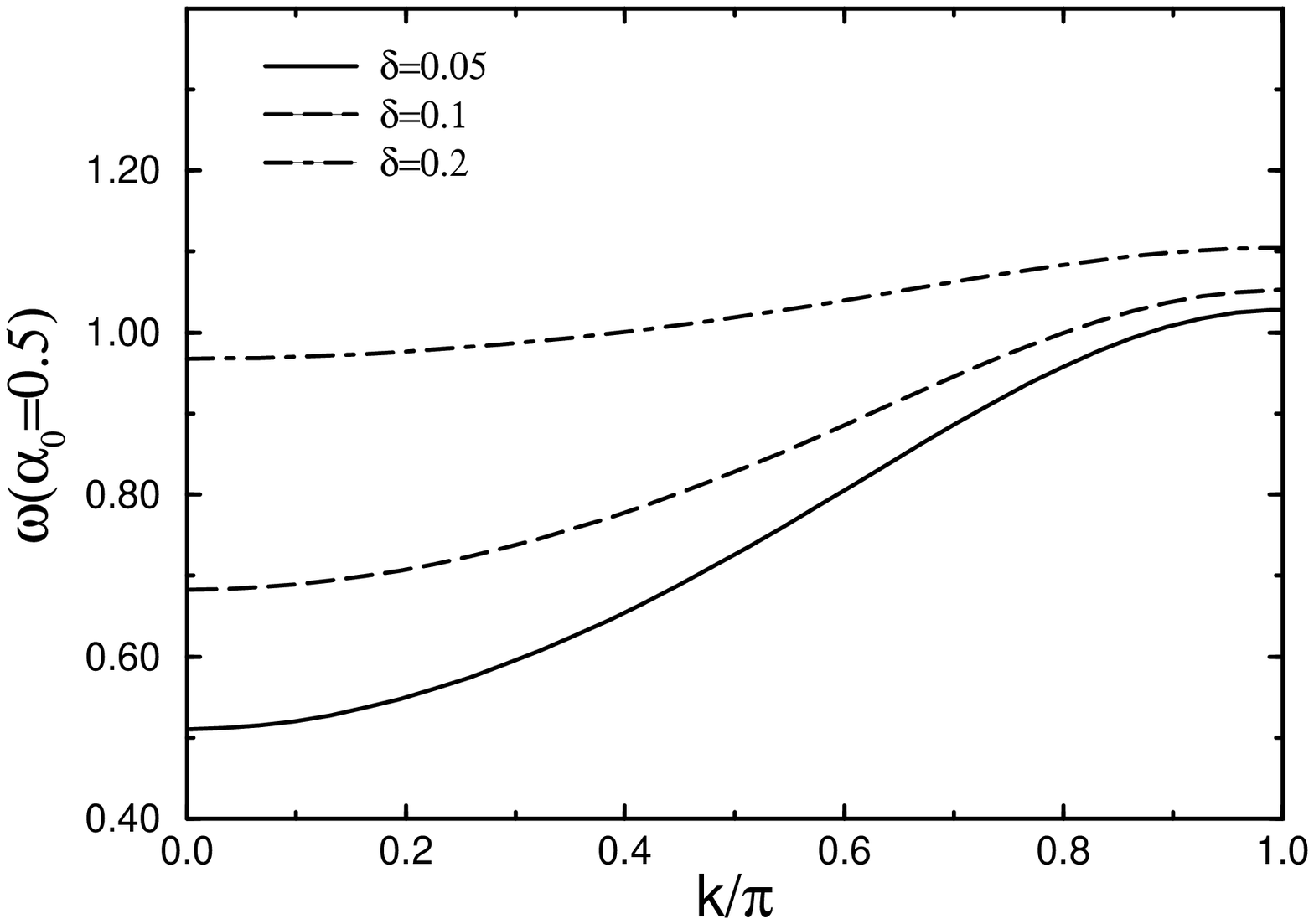}}
\end{picture}
\caption{Dispersion $\omega$ vs. wave vector $k$ at $\alpha_0=0.5$
and various dimerizations. The approach used is the one labeled
}
\label{omega_6_fig}
\end{figure}

So one would like to have a representation at hand which combines the
advantages of the two approaches. A convincing interpolation is achieved
by replacing the Taylor series by the corresponding [5,5]
Pad\'e approximant labeled ``sqrt \& Pad\'e'' in 
Fig.~\ref{omega_5_fig}. Fig.~\ref{omega_6_fig} presents the results of the 
latter approach for finite $\delta$. For completeness,
we mention that  in the $\alpha_0 \le 0.4$ regime
the Pad\'e representation works  as well as the Taylor 
expansion.  For instance, the Pad\'e representation  reproduces the results 
presented in Fig.~\ref{omega_4_fig} within 1\%. This means that for
$\alpha_0 \le 0.4$  there is no need to use the more tedious
Pad\'e representation. But its use for larger frustrations
does not imply an inconsistency of our general approach.
The observation that the dispersion relation for strong frustration
behaves at $k\approx 0$ and at $k\approx \pi$ qualitatively differently
can be understood on physical grounds. For this purpose let us consider
the Majumdar-Ghosh model at $\alpha_0=1/2$ without dimerization. Its 
low-lying excitations are asymptotically  free $S=1/2$ spinons
\cite{shast81a,caspe84,brehm98,mulle99}. 
This implies that in the vicinity of the dispersion
minimum the dispersion relation for $\delta\to 0$ does not represent 
the branch of
a well-defined magnon excitation. But it is the lower band edge of a
two-spinon continuum. In this respect it is similar to the situation
of subcritical frustration. So we have to use the corresponding
 appropriate extrapolation to obtain reliable results.

In the vicinity of the dispersion maximum, however, the spinon interaction
dominates of their kinetic energy and binding occurs even without dimerization.
For instance,
the exact triplet state \cite{caspe82} is a tightly bound two-spinon 
state. A variational analysis shows that this bound triplet exists
below the two-spinon continuum in a finite interval around $k=\pi$
\cite{shast81a,brehm98,mulle99}. Hence, it is fully comprehensible that 
the perturbative approach which starts from local triplets works 
more easily around $k=\pi$ whereas  improved extrapolation is necessary
around $k=0$.

\section{Conclusions}
In summary, we presented a perturbative scheme which relies on 
a suitably chosen unitary transformation.
 The scheme works for an unperturbed equidistant spectrum which can be 
labeled by the number of energy quanta. The perturbation term changes the
number of energy quanta at maximum by a finite number $N$.
Thereby, we generalized the approach by Stein ($N=1$) \cite{stein97}
to general $N$. The unitary transformation
is carried out by flow equations \cite{wegne94}.

By the transformation a systematic mapping of the original problem to
 an effective Hamiltonian is achieved which conserves the number of
energy quanta. Thus Hilbert space sectors with different number of 
energy quanta are separated. Our scheme will be particularly 
useful where ordinary perturbation
theory is hampered by the fact that the structure of the perturbed
states is not known, for instance, systems without translation invariance or
two-particle problems. The possibility to obtain information
in symbolic form, i.e. as polynomials, distinguishes our approach
from the multiprecision method by Barnes {\it et al.} \cite{barne98}
and other  conventional implementations \cite{gelfa90}.

The realization of the perturbative scheme comprises two distinct
steps. The first is still general, the second specific to the model.
For the first step we provided the necessary coefficient up to
order 10 for $N=2$. In written form they are included till order 6
in this publication. The other coefficients shall be provided 
electronically.

The second step is illustrated by dimerized frustrated $S=1/2$ chains.
The limit of isolated dimers has an equidistant spectrum. The ground
state is the product of singlets on the dimers.
The energy quanta are triplets on the dimers. The perturbing weak
couplings between the dimers may create/annihilate at most two triplets
so that $N=2$ holds. After the transformation the number of triplets
is conserved. We presented results for the ground state (zero triplets)
and the magnon dispersion (one triplet). Thereby we demonstrated the
validity and applicability of the scheme proposed.

The results for the spin chains are given as polynomials in the
frustration parameter $\alpha$. So they are easy to use for anybody
who wants to analyze data by appropriate fits. No new calculations
are necessary. The fits can be carried out instantly. Based on 
the results for the dimerization operator $H_{\rm D}$ the 
dependence of the dimerization $\delta$ on the elastic spring constant
$K$ is quickly accessible.

As we demonstrated our results are reliable down to about 6\% 
{\em without} additional information, i.e. using the plain
series. {\em With} additional information as the critical
exponent, for instance, the results can reliably be used down to about
2\%, in some cases even less.
Moreover, starting from the exact coefficients
more elaborate schemes like differential approximants
in two variables become possible \cite{domb89}.

As an outlook we like to point out that our approach can also be used
to compute dispersion relations in two or higher dimensional dimerized
spin systems as demonstrated recently for (VO)$_2$P$_2$O$_7$ \cite{uhrig98c}.
Investigations for CuGeO$_3$ are in preparation. They will improve
 considerably the third order analysis \cite{uhrig97a}.
Another fascinating field concerns computations in the two-magnon sector.
The attractive interaction of two magnons \cite{uhrig96b},
 for instance, leads to bound states. For (VO)$_2$P$_2$O$_7$, 
which is characterized by a relatively large dimerization, such 
results were found in a fourth order calculation.
If the unitary transformation is applied to observables like,
for instance, the Raman operator spectral functions are also
within reach.

\section{Acknowledgements}
We acknowledge many fruitful discussions with E.~M\"uller-Hartmann and 
 F.~Sch\"onfeld. We are indebted to the latter also for the DMRG
results which we used as a benchmark. 
This work was supported by the Deutsche Forschungsgemeinschaft
in the SFB 341 and in the Schwer\-punkt 1073. The large scale computations 
were done on machines of the
 Regional Computing Center of the University of Cologne.


\appendix

\section{Proof of Block Diagonality}
\label{app-proof}
It is shown that the choice (\ref{infgen}) for the infinitesimal generator
achieves block diagonality for the effective Hamiltonian $H(\ell=\infty)$.
The proof follows the lines of Mielkes proof for band matrices \cite{mielk98}.
Let $\{|\nu_i\rangle\}$ be the eigen state basis of $H_0$ and define
\begin{eqnarray}
\nonumber
  h_{ij}(\ell)&:=&\langle\nu_i|H(\ell)|\nu_j\rangle\\
\nonumber
  h_{ij}^0&:=&\langle\nu_i|H_0|\nu_j\rangle\\
\nonumber
  \eta_{ij}(\ell)&:=&\langle\nu_i|\eta(\ell)|\nu_j\rangle\ .
\end{eqnarray}
Inspecting Eqs.~(\ref{hamdef},\ref{thetadef},\ref{infgen}) closely
one realizes that the choice for the infinitesimal generator is
equivalent to
\begin{equation}
\label{ku_gen}
\eta_{ij}(\ell)={\rm sgn}(h^0_{ii}-h^0_{ij})h_{ij}(\ell)\ .
\end{equation}
Inserting this expression in flow equation (\ref{flow}) yields
\begin{eqnarray}
  \nonumber
  \lefteqn{\frac{\partial h_{ij}}{\partial \ell} = -{\rm
  sgn}(h^0_{ii}-h^0_{jj})(h_{ii}-h_{jj})h_{ij} }
\\
\label{dia_summ_i}
& &\mbox{} + \sum_{k\neq
  i,j}({\rm sgn}(h^0_{ii}-h^0_{kk})+{\rm
  sgn}(h^0_{jj}-h^0_{kk}))|h_{ik}|^2 \ .
\end{eqnarray}
Assume without loss of generality
 that the eigen states $|\nu_i\rangle$ are labeled such that
$h_{kk}^0\ge h_{ii}^0$ if $k>i$.
Let us consider the sum of the first $r$ diagonal elements of $H(\ell)$
\begin{equation}
\label{abl}
\frac{\partial}{\partial \ell}\sum_{i=1}^r
  h_{ii}=2\sum_{i=1}^r\sum_{k>r}{\rm
  sgn}(h^0_{ii}-h^0_{kk})|h_{ik}|^2 \ .
\end{equation}
The right side of Eq.~(\ref{abl}) is non-positive. 
Thus the sum on the left hand side
is a continuous monotonically decreasing function with $\ell$.
If we know that it is bounded from below we conclude that the sum
converges whence the vanishing  of its derivative for $\ell\to\infty$
ensues immediately. Hence we need beyond the conditions (i) and (ii)
in the Introduction the boundedness for the whole Hamiltonian.
If this is not given our choice for $\eta$ might be problematic
whereas Wegner's choice still works (for an example, see Ref. \cite{creme99}).

As we are, however, interested in deriving a perturbation expansion
order by order we can assume the whole Hamiltonian to be bounded from below
without loss of generality. For any finite order of the expansion
it is sufficient to consider a finite cluster supposing some short range
interaction.
Then the Hamiltonian is a finite dimensional matrix and is
certainly bounded from below.  This is true in particular
if we stay on the abstract level as in Eq.~(\ref{outset1}).
The generalized variational principle implies that
\begin{equation}
\sum_{i=1}^r  h_{ii} \ge \sum_{i=1}^r \omega_i
\end{equation}
holds for all $\ell$ where the $\omega_i$ are the eigen values
of $H$ in ascending order. Note that the eigen values are invariant
under the unitary transformation.

The vanishing of the derivative of $\sum_{i=1}^r
  h_{ii}$ for arbitrary $r$ implies eventually
\begin{equation}
\lim_{\ell\to\infty}{\rm
  sgn}(h^0_{ii}-h^0_{kk})|h_{ik}|^2 = 0\ .
\end{equation}
From this equation follows that {\em either} the eigen states $|\nu_i\rangle$
and $|\nu_k\rangle$ are degenerate, i.e.~they belong to the same block,
{\em or} $h_{ik}(\ell=\infty)=0$, i.e.~matrix elements linking different
blocks vanish. Hence block diagonality is achieved and the number of energy 
quanta given by $H_0$ is conserved
\begin{equation}
\lim_{\ell \rightarrow \infty}[H_0,H(\ell)]=0\ .
\end{equation}
This concludes the formal proof. 

To restrict the argument at one stage to finite clusters does not
constitute a real restriction for the series expansion in any finite
order. The linked cluster theorem tells us that any finite order
can be found from an appropriate finite cluster. 

In practice, the
important issues is whether the physics remains the same on variation
of the expansion parameter $\lambda$. In our example of dimerized
chains the expansion makes sense as long as the gap does not close.
In other words, those situations are accessible which can be linked
continuously to the dimer limit by gapped systems. Gapless
situations can only be described if they are the limit of gapped systems
linked to the dimer limit.

\onecolumn
\section{Coefficients of the Effective Hamiltonian}
\label{app-coeff}
In the following tables the coefficients $C(\underline{m})$ of the 
effective Hamiltonian (\ref{effhamilfin}) are given up to sixth
order inclusively. Order 7 to 10 will be available electronically.\\[2mm]
\begin{tabular}[t]{|cc|}
\hline
$\underline{m}$ & $C(\underline{m})$\\
\hline\hline
$|\underline{m}|=1$&\\
\hline
0 &  1\\
\hline
$|\underline{m}|=2$&\\
\hline
1-1 &  1\\
2-2 &  1/2\\
\hline 
$|\underline{m}|=3$&\\
\hline
01-1 &  -1/2\\
02-2 &  -1/8\\
10-1 &  1\\
11-2 &  1/2\\
1-21 &  -1\\
20-2 &  1/4\\
\hline
$|\underline{m}|=4$&\\
\hline
001-1 & 1/4\\
002-2 & 1/32\\
010-1 & -1\\
011-2 & -3/8\\
01-21 & 1/4\\
01-10 & 1/2\\
020-2 & -1/8\\
02-20 & 1/16\\
02-1-1 & -3/8\\
100-1 & 1\\
101-2 & 1/2\\
10-21 & -1\\
110-2 & 1/4\\
11-1-1 & 1/2\\
12-2-1 & 1/3\\
12-1-2 & 1/6\\
1-22-1 & -1\\
1-2-12 & 1/2\\
1-11-1 & -1\\
1-12-2 & -3/8\\
1-1-22 & 1/8\\
200-2 & 1/8\\
21-1-2 & 1/12\\
22-2-2 & 1/16\\
2-22-2 & -1/8\\
2-11-2 & 1/4\\
\hline
$|\underline{m}|=5$&\\
\hline
0001-1 &  -1/8\\
0002-2 &  -1/128\\
0010-1 &  3/4\\
0011-2 &  7/32\\
001-21 &  -5/16\\
001-10 &  -3/8\\
0020-2 &  3/64\\
002-20 &  -3/128\\
002-1-1 &  7/32\\
0100-1 &  -3/2\\
0101-2 &  -5/8\\
\hline
\end{tabular}
\begin{tabular}[t]{|cc|}
\hline
$\underline{m}$ & $C(\underline{m})$\\
\hline\hline
$|\underline{m}|=5$&\\
\hline
010-21 &  7/8\\
010-10 &  3/2\\
0110-2 &  -1/4\\
011-20 &  7/16\\
011-1-1 &  -5/8\\
012-2-1 &  -7/18\\
012-1-2 &  -11/72\\
01-201 &  1/8\\
01-210 &  -3/8\\
01-22-1 &  1/2\\
01-2-12 &  -1/24\\
01-11-1 &  9/8\\
01-12-2 &  35/96\\
01-1-22 &  -19/96\\
01-1-11 &  -3/8\\
0200-2 &  -3/32\\
020-20 &  3/32\\
020-1-1 &  -1/4\\
021-2-1 &  -11/72\\
021-1-2 &  -1/18\\
022-2-2 &  -5/128\\
02-21-1 &  7/24\\
02-22-2 &  9/128\\
02-2-22 &  -3/128\\
02-2-11 &  -1/12\\
02-10-1 &  -5/8\\
02-11-2 &  -1/4\\
02-1-21 &  7/24\\
1000-1 &  1\\
1001-2 &  1/2\\
100-21 &  -1\\
1010-2 &  1/4\\
101-1-1 &  1/2\\
102-2-1 &  1/3\\
102-1-2 &  1/6\\
10-201 &  1\\
10-22-1 &  -1\\
10-2-12 &  1/2\\
10-11-1 &  -3/2\\
10-12-2 &  -5/8\\
10-1-22 &  3/8\\
10-1-11 &  1/2\\
1100-2 &  1/8\\
110-1-1 &  1/4\\
111-2-1 &  1/6\\
111-1-2 &  1/12\\
112-2-2 &  1/16\\
11-21-1 &  -5/8\\
11-22-2 &  -1/4\\
11-2-22 &  1/8\\
\hline
\end{tabular}
\begin{tabular}[t]{|cc|}
\hline
$\underline{m}$ & $C(\underline{m})$\\
\hline\hline
$|\underline{m}|=5$&\\
\hline
11-2-11 &  1/8\\
11-11-2 &  1/4\\
11-1-21 &  -1/2\\
120-2-1 &  1/9\\
120-1-2 &  1/18\\
121-2-2 &  1/24\\
12-21-2 &  1/6\\
12-2-21 &  -1/3\\
12-10-2 &  1/12\\
1-202-1 &  1\\
1-20-12 &  -1/2\\
1-211-1 &  1/4\\
1-212-2 &  0\\
1-21-22 &  1/4\\
1-21-11 &  3/4\\
1-221-2 &  -1/2\\
1-22-21 &  1\\
1-2-212 &  -1/6\\
1-2-102 &  -1/4\\
1-101-1 &  3/4\\
1-102-2 &  7/32\\
1-10-22 &  -3/32\\
1-10-11 &  -1/4\\
1-111-2 &  -5/8\\
1-120-2 &  -1/4\\
1-1-202 &  0\\
1-1-1-12 &  1/8\\
2000-2 &  1/16\\
201-1-2 &  1/24\\
202-2-2 &  1/32\\
20-22-2 &  -3/32\\
20-2-22 &  1/32\\
20-11-2 &  1/8\\
210-1-2 &  1/36\\
211-2-2 &  1/48\\
21-21-2 &  1/12\\
220-2-2 &  1/64\\
2-202-2 &  3/64\\
2-20-22 &  -1/64\\
2-211-2 &  -1/4\\
2-2-1-12 &  1/8\\
2-101-2 &  1/4\\
2-1-2-12 &  1/4\\
\hline
$|\underline{m}|=6$&\\
\hline
00001-1 &  1/16\\
00002-2 &  1/512\\
00010-1 &  -1/2\\
00011-2 &  -15/128\\
0001-21 &  5/64\\
0001-10 &  1/4\\
\hline
\end{tabular}
\begin{tabular}[t]{|cc|}
\hline
$\underline{m}$ & $C(\underline{m})$\\
\hline\hline
$|\underline{m}|=6$&\\
\hline
00020-2 &  -1/64\\
0002-20 &  1/128\\
0002-1-1 &  -15/128\\
00100-1 &  3/2\\
00101-2 &  17/32\\
0010-21 &  -3/4\\
0010-10 &  -3/2\\
00110-2 &  11/64\\
0011-20 &  -45/128\\
0011-1-1 &  17/32\\
0012-2-1 &  17/54\\
0012-1-2 &  85/864\\
001-201 &  1/2\\
001-210 &  1/64\\
001-22-1 &  -13/24\\
001-2-12 &  53/288\\
001-100 &  3/8\\
001-11-1 &  -29/32\\
001-12-2 &  -301/1152\\
001-1-22 &  179/1152\\
001-1-11 &  7/32\\
00200-2 &  3/64\\
0020-20 &  -3/64\\
0020-1-1 &  11/64\\
0021-2-1 &  85/864\\
0021-1-2 &  43/1728\\
0022-2-2 &  17/1024\\
002-200 &  3/256\\
002-21-1 &  -197/1152\\
002-22-2 &  -29/1024\\
002-2-22 &  7/1024\\
002-2-11 &  -23/1152\\
002-10-1 &  17/32\\
002-11-2 &  11/64\\
002-1-21 &  -65/288\\
002-1-10 &  -45/128\\
01000-1 &  -2\\
01001-2 &  -7/8\\
0100-21 &  23/16\\
0100-10 &  3\\
01010-2 &  -3/8\\
0101-20 &  17/16\\
0101-1-1 &  -7/8\\
0102-2-1 &  -5/9\\
0102-1-2 &  -17/72\\
010-201 &  -5/8\\
010-210 &  -3/4\\
010-22-1 &  9/8\\
010-2-12 &  -7/18\\
010-11-1 &  39/16\\
\hline
\end{tabular}
\begin{tabular}[t]{|cc|}
\hline
$\underline{m}$ & $C(\underline{m})$\\
\hline\hline
$|\underline{m}|=6$&\\
\hline
010-12-2 &  269/288\\
010-1-22 &  -181/288\\
010-1-11 &  -17/16\\
01100-2 &  -5/32\\
0110-20 &  11/32\\
0110-1-1 &  -3/8\\
0111-2-1 &  -17/72\\
0111-1-2 &  -7/72\\
0112-2-2 &  -9/128\\
011-21-1 &  5/6\\
011-22-2 &  113/384\\
011-2-22 &  -67/384\\
011-2-11 &  -7/24\\
011-10-1 &  -7/8\\
011-11-2 &  -3/8\\
011-1-21 &  7/12\\
011-1-10 &  17/16\\
0120-2-1 &  -4/27\\
0120-1-2 &  -13/216\\
0121-2-2 &  -25/576\\
012-20-1 &  -5/9\\
012-21-2 &  -17/72\\
012-2-21 &  103/288\\
012-2-10 &  17/27\\
012-10-2 &  -7/72\\
012-1-20 &  85/432\\
012-1-1-1 &  -17/72\\
01-2001 &  -9/16\\
01-202-1 &  0\\
01-20-12 &  -1/6\\
01-211-1 &  -2/3\\
01-212-2 &  -15/64\\
01-21-22 &  37/192\\
01-21-11 &  13/24\\
01-220-1 &  7/8\\
01-221-2 &  1/3\\
01-22-21 &  -5/16\\
01-22-10 &  -11/12\\
01-2-212 &  -1/144\\
01-2-221 &  -25/288\\
01-2-102 &  -1/36\\
01-2-111 &  -1/6\\
01-2-120 &  31/144\\
01-101-1 &  -1\\
01-102-2 &  -329/1152\\
01-10-22 &  53/384\\
01-10-11 &  3/16\\
01-110-1 &  33/16\\
01-111-2 &  73/96\\
01-11-21 &  -49/48\\
\hline
\end{tabular}
\begin{tabular}[t]{|cc|}
\hline
$\underline{m}$ & $C(\underline{m})$\\
\hline\hline
$|\underline{m}|=6$&\\
\hline
01-11-10 &  -27/16\\
01-120-2 &  151/576\\
01-12-20 &  -149/384\\
01-12-1-1 &  73/96\\
01-1-202 &  47/576\\
01-1-211 &  17/96\\
01-1-220 &  113/1152\\
01-1-101 &  7/16\\
01-1-12-1 &  -31/48\\
01-1-1-12 &  17/96\\
02000-2 &  -1/16\\
0200-20 &  3/32\\
0200-1-1 &  -5/32\\
0201-2-1 &  -7/72\\
0201-1-2 &  -11/288\\
0202-2-2 &  -7/256\\
020-21-1 &  73/288\\
020-22-2 &  39/512\\
020-2-22 &  -17/512\\
020-2-11 &  -17/288\\
020-10-1 &  -3/8\\
020-11-2 &  -5/32\\
020-1-21 &  2/9\\
0210-2-1 &  -13/216\\
0210-1-2 &  -5/216\\
0211-2-2 &  -19/1152\\
021-20-1 &  -17/72\\
021-21-2 &  -7/72\\
021-2-21 &  19/144\\
021-10-2 &  -11/288\\
021-1-20 &  43/864\\
021-1-1-1 &  -7/72\\
0220-2-2 &  -3/256\\
022-20-2 &  -7/256\\
022-2-20 &  17/512\\
022-2-1-1 &  -9/128\\
022-1-2-1 &  -25/576\\
022-1-1-2 &  -19/1152\\
02-201-1 &  -211/1152\\
02-202-2 &  -1/32\\
02-20-22 &  3/512\\
02-20-11 &  -7/384\\
02-210-1 &  95/144\\
02-211-2 &  85/384\\
02-21-21 &  -53/192\\
02-220-2 &  33/512\\
02-22-20 &  -27/512\\
02-22-1-1 &  85/384\\
02-2-202 &  7/512\\
02-2-211 &  23/384\\
\hline
\end{tabular}
\begin{tabular}[t]{|cc|}
\hline
$\underline{m}$ & $C(\underline{m})$\\
\hline\hline
$|\underline{m}|=6$&\\
\hline
02-2-101 &  31/144\\
02-2-12-1 &  -11/64\\
02-2-1-12 &  23/384\\
02-100-1 &  -7/8\\
02-101-2 &  -3/8\\
02-10-21 &  7/12\\
02-110-2 &  -5/32\\
02-11-20 &  11/32\\
02-11-1-1 &  -3/8\\
02-12-2-1 &  -17/72\\
02-12-1-2 &  -7/72\\
02-1-201 &  -5/36\\
02-1-22-1 &  5/12\\
02-1-2-12 &  -11/96\\
02-1-11-1 &  5/6\\
02-1-12-2 &  113/384\\
02-1-1-22 &  -67/384\\
02-1-1-11 &  -7/24\\
10000-1 &  1\\
10001-2 &  1/2\\
1000-21 &  -1\\
10010-2 &  1/4\\
1001-1-1 &  1/2\\
1002-2-1 &  1/3\\
1002-1-2 &  1/6\\
100-201 &  1\\
100-22-1 &  -1\\
100-2-12 &  1/2\\
100-11-1 &  -2\\
100-12-2 &  -7/8\\
100-1-22 &  5/8\\
100-1-11 &  1\\
10100-2 &  1/8\\
1010-1-1 &  1/4\\
1011-2-1 &  1/6\\
1011-1-2 &  1/12\\
1012-2-2 &  1/16\\
101-21-1 &  -7/8\\
101-22-2 &  -3/8\\
101-2-22 &  1/4\\
101-2-11 &  3/8\\
101-10-1 &  1/2\\
101-11-2 &  1/4\\
101-1-21 &  -1/2\\
1020-2-1 &  1/9\\
1020-1-2 &  1/18\\
1021-2-2 &  1/24\\
102-20-1 &  1/3\\
102-21-2 &  1/6\\
102-2-21 &  -1/3\\
102-10-2 &  1/12\\
102-1-1-1 &  1/6\\
10-202-1 &  1\\
10-20-12 &  -1/2\\
10-211-1 &  5/8\\
\hline
\end{tabular}
\begin{tabular}[t]{|cc|}
\hline
$\underline{m}$ & $C(\underline{m})$\\
\hline\hline
$|\underline{m}|=6$&\\
\hline
10-212-2 &  3/16\\
10-21-22 &  1/16\\
10-21-11 &  3/8\\
10-220-1 &  -1\\
10-221-2 &  -1/2\\
10-22-21 &  1\\
10-2-212 &  -1/6\\
10-2-221 &  -1/3\\
10-2-102 &  -1/4\\
10-2-111 &  -1/2\\
10-101-1 &  3/2\\
10-102-2 &  17/32\\
10-10-22 &  -9/32\\
10-10-11 &  -1/2\\
10-110-1 &  -2\\
10-111-2 &  -7/8\\
10-11-21 &  5/4\\
10-120-2 &  -3/8\\
10-12-1-1 &  -7/8\\
10-1-202 &  -1/8\\
10-1-211 &  -1/8\\
10-1-12-1 &  3/4\\
10-1-1-12 &  -1/8\\
11000-2 &  1/16\\
1100-1-1 &  1/8\\
1101-2-1 &  1/12\\
1101-1-2 &  1/24\\
1102-2-2 &  1/32\\
110-21-1 &  -3/8\\
110-22-2 &  -5/32\\
110-2-22 &  3/32\\
110-2-11 &  1/8\\
110-11-2 &  1/8\\
110-1-21 &  -1/4\\
1110-2-1 &  1/18\\
1110-1-2 &  1/36\\
1111-2-2 &  1/48\\
111-21-2 &  1/12\\
111-2-21 &  -1/6\\
111-10-2 &  1/24\\
111-1-1-1 &  1/12\\
1120-2-2 &  1/64\\
112-20-2 &  1/32\\
112-2-1-1 &  1/16\\
112-1-2-1 &  1/24\\
112-1-1-2 &  1/48\\
11-201-1 &  17/32\\
11-202-2 &  11/64\\
11-20-22 &  -5/64\\
11-20-11 &  -5/32\\
11-211-2 &  -3/8\\
11-21-21 &  1/2\\
11-220-2 &  -5/32\\
11-22-1-1 &  -3/8\\
11-2-202 &  -1/32\\
\hline
\end{tabular}
\begin{tabular}[t]{|cc|}
\hline
$\underline{m}$ & $C(\underline{m})$\\
\hline\hline
$|\underline{m}|=6$&\\
\hline
11-2-12-1 &  1/4\\
11-101-2 &  1/4\\
11-10-21 &  -1/2\\
11-110-2 &  1/8\\
11-11-1-1 &  1/4\\
11-12-2-1 &  1/6\\
11-12-1-2 &  1/12\\
11-1-22-1 &  -1/2\\
11-1-2-12 &  1/4\\
11-1-11-1 &  -7/8\\
11-1-12-2 &  -3/8\\
11-1-1-22 &  1/4\\
11-1-1-11 &  3/8\\
1200-2-1 &  1/27\\
1200-1-2 &  1/54\\
1201-2-2 &  1/72\\
120-21-2 &  1/18\\
120-2-21 &  -1/9\\
120-10-2 &  1/36\\
1210-2-2 &  1/96\\
121-20-2 &  1/48\\
121-1-2-1 &  1/36\\
121-1-1-2 &  1/72\\
122-2-2-1 &  1/45\\
122-2-1-2 &  1/90\\
122-1-2-2 &  1/120\\
12-201-2 &  1/6\\
12-20-21 &  -1/3\\
12-210-2 &  1/12\\
12-22-2-1 &  1/9\\
12-22-1-2 &  1/18\\
12-2-22-1 &  -1/3\\
12-2-2-12 &  1/6\\
12-2-11-1 &  -5/9\\
12-2-12-2 &  -17/72\\
12-2-1-22 &  11/72\\
12-2-1-11 &  2/9\\
12-100-2 &  1/24\\
12-11-2-1 &  1/18\\
12-11-1-2 &  1/36\\
12-12-2-2 &  1/48\\
12-1-21-1 &  -17/72\\
12-1-22-2 &  -7/72\\
12-1-2-22 &  1/18\\
12-1-2-11 &  5/72\\
12-1-11-2 &  1/12\\
12-1-1-21 &  -1/6\\
1-2002-1 &  -1\\
1-200-12 &  1/2\\
1-2011-1 &  3/8\\
1-2012-2 &  5/16\\
1-201-22 &  -9/16\\
1-201-11 &  -11/8\\
1-2021-2 &  1/2\\
1-202-21 &  -1\\
\hline
\end{tabular}
\begin{tabular}[t]{|cc|}
\hline
$\underline{m}$ & $C(\underline{m})$\\
\hline\hline
$|\underline{m}|=6$&\\
\hline
1-20-212 &  1/6\\
1-20-102 &  1/4\\
1-2101-1 &  -7/16\\
1-2102-2 &  -5/32\\
1-210-22 &  7/32\\
1-210-11 &  11/16\\
1-2111-2 &  1/4\\
1-2120-2 &  1/16\\
1-21-202 &  -3/16\\
1-21-12-1 &  1/2\\
1-21-1-12 &  -1/2\\
1-2201-2 &  -1/2\\
1-2210-2 &  -1/4\\
1-222-1-2 &  -1/6\\
1-22-22-1 &  1\\
1-22-2-12 &  -1/2\\
1-22-11-1 &  1\\
1-22-12-2 &  3/8\\
1-22-1-22 &  -1/8\\
1-22-1-11 &  0\\
1-2-2012 &  1/18\\
1-2-2102 &  1/12\\
1-2-222-1 &  -1/3\\
1-2-22-12 &  1/6\\
1-2-2-122 &  1/24\\
1-2-1002 &  1/8\\
1-2-112-1 &  -1/2\\
1-2-11-12 &  1/4\\
1-2-121-1 &  1/24\\
1-2-122-2 &  1/12\\
1-2-12-22 &  -5/24\\
1-2-12-11 &  -13/24\\
1-2-1-222 &  1/16\\
1-2-1-112 &  1/12\\
1-1001-1 &  -1/2\\
1-1002-2 &  -15/128\\
1-100-22 &  5/128\\
1-1011-2 &  17/32\\
1-1020-2 &  11/64\\
1-10-202 &  3/64\\
1-10-1-12 &  5/32\\
1-1101-2 &  -7/8\\
1-1110-2 &  -3/8\\
1-112-1-2 &  -17/72\\
1-11-2-12 &  -7/24\\
1-11-11-1 &  2\\
1-11-12-2 &  71/96\\
1-11-1-22 &  -43/96\\
1-11-1-11 &  -3/4\\
1-1200-2 &  -5/32\\
1-121-1-2 &  -7/72\\
1-122-2-2 &  -9/128\\
1-12-21-1 &  31/48\\
1-12-22-2 &  83/384\\
1-12-2-22 &  -41/384\\
\hline
\end{tabular}
\begin{tabular}[t]{|cc|}
\hline
$\underline{m}$ & $C(\underline{m})$\\
\hline\hline
$|\underline{m}|=6$&\\
\hline
1-12-2-11 &  -3/16\\
1-12-11-2 &  -3/8\\
1-1-2002 &  -1/32\\
1-1-21-12 &  -1/8\\
1-1-221-1 &  -11/48\\
1-1-222-2 &  -25/384\\
1-1-22-22 &  19/384\\
1-1-2-222 &  -1/128\\
1-1-2-112 &  -1/72\\
1-1-10-12 &  -3/8\\
1-1-111-1 &  -1/2\\
1-1-112-2 &  -19/96\\
1-1-11-22 &  23/96\\
1-1-121-2 &  5/24\\
1-1-1-212 &  -5/72\\
1-1-1-102 &  -1/8\\
20000-2 &  1/32\\
2001-1-2 &  1/48\\
2002-2-2 &  1/64\\
200-22-2 &  -1/16\\
200-2-22 &  1/32\\
200-11-2 &  1/16\\
2010-1-2 &  1/72\\
2011-2-2 &  1/96\\
201-21-2 &  1/24\\
201-10-2 &  1/48\\
2020-2-2 &  1/128\\
202-20-2 &  1/64\\
202-1-1-2 &  1/96\\
20-202-2 &  3/64\\
20-20-22 &  -1/64\\
20-211-2 &  -5/32\\
20-220-2 &  -1/16\\
20-2-1-12 &  1/32\\
20-101-2 &  1/8\\
20-110-2 &  1/16\\
20-12-1-2 &  1/24\\
20-1-2-12 &  1/8\\
20-1-12-2 &  -5/32\\
20-1-1-22 &  3/32\\
2100-1-2 &  1/108\\
2101-2-2 &  1/144\\
210-21-2 &  1/36\\
2110-2-2 &  1/192\\
211-1-1-2 &  1/144\\
212-2-1-2 &  1/180\\
212-1-2-2 &  1/240\\
21-201-2 &  1/12\\
21-22-1-2 &  1/36\\
21-2-2-12 &  1/12\\
21-2-12-2 &  -7/72\\
21-2-1-22 &  1/18\\
21-11-1-2 &  1/72\\
21-12-2-2 &  1/96\\
21-1-22-2 &  -11/288\\
\hline
\end{tabular}
\begin{tabular}[t]{|cc|}
\hline
$\underline{m}$ & $C(\underline{m})$\\
\hline\hline
$|\underline{m}|=6$&\\
\hline
21-1-2-22 &  5/288\\
21-1-11-2 &  1/24\\
2200-2-2 &  1/256\\
221-1-2-2 &  1/320\\
222-2-2-2 &  1/384\\
22-22-2-2 &  1/128\\
22-2-22-2 &  -7/256\\
22-2-2-22 &  3/256\\
22-2-11-2 &  1/32\\
22-11-2-2 &  1/192\\
22-1-21-2 &  1/48\\
2-2002-2 &  -1/64\\
2-2011-2 &  11/64\\
2-20-1-12 &  5/64\\
2-2101-2 &  -3/8\\
2-21-2-12 &  -1/12\\
2-21-12-2 &  49/192\\
2-21-1-22 &  -9/64\\
2-22-22-2 &  1/16\\
2-22-2-22 &  -3/128\\
2-22-11-2 &  -5/32\\
2-2-21-12 &  -3/32\\
2-2-222-2 &  -1/64\\
2-2-10-12 &  -1/4\\
2-2-112-2 &  -17/192\\
2-2-121-2 &  1/24\\
2-1001-2 &  1/4\\
2-10-2-12 &  1/4\\
2-11-11-2 &  1/8\\
2-12-21-2 &  1/12\\
2-1-221-2 &  -1/4\\
2-1-111-2 &  -3/8\\
\hline
\end{tabular}

\section{Effective Hopping Elements for the Spin Chain}
\label{app-effhopp}
In the following the effective hopping elements as they
appear in Eq.~(\ref{hopp_coef}) are given up to order 6. 
The effective hopping elements up to order 10  will be provided electronically.
We substituted $\overline{\alpha} = 1 - 2\alpha$ and 
$\overline{\lambda} = \frac{1}{4}\lambda$.
\begin{eqnarray*}
a_0-E_0 &=& 1-\bigg(4-3\overline{\alpha}^{2}\bigg)\overline{\lambda}^{2} 
- \bigg(8-8\overline{\alpha} -
6\overline{\alpha}^{2} + 3\overline{\alpha}^{
3}\bigg)\overline{\lambda} ^{3} 
-\bigg(2 - 24\overline{\alpha} +
5\overline{\alpha}^{2} + 8\overline{\alpha}^{3} + {\displaystyle
\frac {13}{4}} \overline{\alpha}^{4} \bigg)\overline{\lambda} ^{4}
 + \bigg(56 - 82\overline{\alpha} - 22\overline{\alpha}^{2}
\\
& & \mbox{}  + 55\overline{\alpha}^{3} - 39\overline{\alpha}^{4} + 20
\overline{\alpha}^{5}\bigg)\overline{\lambda} ^{5}
 + \bigg( {\displaystyle 
\frac {367}{3}} - {\displaystyle \frac {7328}{9}} \overline{\alpha} +
{\displaystyle \frac {22976}{27}} \overline{\alpha}^{2} + {\displaystyle 
\frac {6442}{27}} \overline{\alpha}^{3} -{\displaystyle \frac {28895}{54}} 
\overline{\alpha}^{4} + 193\overline{\alpha}^{5} - 32\overline{\alpha}^{6}
 \bigg)\overline{\lambda} ^{6} 
\end{eqnarray*}

\begin{eqnarray*}
a_1 &=& - 2\overline{\alpha}\overline{\lambda}  - 
4\overline{\lambda} ^{2} - \bigg(8-8\overline{\alpha} - 2\overline{\alpha}^{3}
\bigg)\overline{\lambda} ^{3} + \bigg(4+20\overline{\alpha} - 
24\overline{\alpha}^{2} + 10\overline{\alpha}^{3} - 5\overline{\alpha}^{4}
 \bigg)\overline{\lambda} ^{4} 
+\bigg(92- {\displaystyle \frac {499}{3}} \overline{\alpha} -
{\displaystyle \frac {164}{3}} \overline{\alpha}^{2} + 
152\overline{\alpha}^{3}
\\
& & \mbox{} 
 - 47\overline{\alpha}^{4}
 + {\displaystyle \frac {13}{2}} \overline{\alpha}^{5}
\bigg)\overline{\lambda} ^{5}
+ \bigg( {\displaystyle 
\frac {532}{3}} - {\displaystyle \frac {11906}{9}} \overline{\alpha} + 
{\displaystyle \frac {11960}{9}} \overline{\alpha}^{2} + {\displaystyle \frac {
1648}{3}} \overline{\alpha}^{3} - {\displaystyle \frac {41357}{54}} 
\overline{\alpha}^{4} +
85\overline{\alpha}^{5} + 6\overline{\alpha}^{6}\bigg)\overline{\lambda} 
^{6}
\end{eqnarray*}

\begin{eqnarray*}
a_2 &=& - \overline{\alpha}^{2}\overline{\lambda}^{2}-
\bigg(4\overline{\alpha}^{2} - 2\overline{\alpha}^{3}\bigg)
\overline{\lambda} ^{3}  + \bigg(6- 4\overline{\alpha} -
23\overline{\alpha}^{2} + 14\overline{\alpha}^{3} - 
{\displaystyle \frac {1}{2}} \overline{\alpha}^{4}\bigg)\overline{\lambda}^{4}
 + \bigg(36 - {\displaystyle \frac {272}{3}} \overline{\alpha} -
 {\displaystyle \frac {220}{3}} \overline{\alpha}^{2} + 
{\displaystyle \frac {1150}{9}} \overline{\alpha}^{3} - 9\overline{\alpha}^{4}
\\  
& & \mbox{}
- {\displaystyle \frac {13}{2}} \overline{\alpha}^{5}\bigg)
\overline{\lambda} ^{5}  + \bigg({\displaystyle 
\frac {107}{3}} - {\displaystyle \frac {1630}{3}} \overline{\alpha} +
{\displaystyle \frac {1126}{3}} \overline{\alpha}^{2} + {\displaystyle \frac {
5102}{9}} \overline{\alpha}^{3} - {\displaystyle \frac {13205}{36}} 
\overline{\alpha}^{4} - 59\overline{\alpha}^{5} + 11\overline{\alpha}^{6}
\bigg)\overline{\lambda}^{6}
\end{eqnarray*}

\begin{eqnarray*}
a_3 &=&- \overline{\alpha}^{3}\overline{\lambda} ^{3} -
\bigg({\displaystyle \frac {10}{3}} \overline{\alpha}^{2} + 
4\overline{\alpha}^{3} - 2
\overline{\alpha}^{4}\bigg)\overline{\lambda} ^{4} - 
\bigg( {\displaystyle \frac {19}{3}} \overline{\alpha} +
20\overline{\alpha}^{2} - {\displaystyle \frac {10}{3}} \overline{\alpha}^{3}
 - 11\overline{\alpha}^{4} - 3
\overline{\alpha}^{5}\bigg)\overline{\lambda} ^{5}
- \bigg( 
{\displaystyle \frac {58}{3}}+ {\displaystyle \frac {104}{3}} 
\overline{\alpha} -
{\displaystyle \frac {224}{9}} \overline{\alpha}^{2} -63\overline{\alpha}^{3} 
\\
& & \mbox{}
 + 
{\displaystyle \frac {103}{2}} \overline{\alpha}^{4} - {\displaystyle \frac {57
}{2}} \overline{\alpha}^{5} + {\displaystyle \frac {81}{4}} 
\overline{\alpha}^{6} \bigg)\overline{\lambda} ^{6} 
\end{eqnarray*}

\begin{eqnarray*}
a_4 &=&- {\displaystyle \frac {5}{4}} \overline{\alpha}^{4}
\overline{\lambda} 
^{4}-\bigg( {\displaystyle \frac {40}{9}} \overline{\alpha}^{3} + 
6\overline{\alpha}^{4} - 3
\overline{\alpha}^{5}\bigg)\overline{\lambda} ^{5} + \bigg({\displaystyle 
\frac {11}{3}} \overline{\alpha}^{2}
 - {\displaystyle \frac {827}{27}} \overline{\alpha}^{3} - {\displaystyle 
\frac {1127}{36}} \overline{\alpha}^{4} + {\displaystyle \frac {91}{4}} 
\overline{\alpha}^{5}
 + {\displaystyle \frac {73}{16}} \overline{\alpha}^{6}\bigg)
\overline{\lambda} ^{6}
\end{eqnarray*}

\begin{eqnarray*}
a_5&=&- {\displaystyle \frac {7}{4}} \overline{\alpha}^{5}
\overline{\lambda} ^{5}-\bigg( {\displaystyle \frac {497}{54}} 
\overline{\alpha}^{4} + 10\overline{\alpha}^{5}
 - 5\overline{\alpha}^{6}\bigg)\overline{\lambda} ^{6}
\qquad\qquad
a_6 \;=\; - {\displaystyle \frac {21}{8}} \overline{\alpha}^{6}
\overline{\lambda} ^{6}
\end{eqnarray*}

\end{document}